\documentclass[12pt]{article}
\pdfoutput=1

\def \bea { \begin{eqnarray}}
\def \eea {\end{eqnarray}}
\def \be {\begin{equation}}
\def \ee {\end{equation}}

\usepackage{type1cm,amsmath,amssymb,color,graphics,amscd,amsfonts,mathrsfs,epsf,indentfirst}
\usepackage{bm}
\usepackage{subfigure}
\usepackage{multirow,graphicx}
\usepackage{epstopdf}

\title{Stable non-uniform black strings below the critical dimension}
\author{Pau Figueras$^1$, Keiju Murata$^{1,2}$ and Harvey S. Reall$^1$
\\ 
{\small 
$^1$DAMTP, Centre for Mathematical Sciences, Wilberforce Road,
Cambridge CB3 0WA, UK}
\\
{\small
$^2$Yukawa Institute for Theoretical Physics, Kyoto University, Kyoto, 606-8502,
Japan
}
}

\begin{document}
\maketitle

\begin{abstract}
The higher-dimensional vacuum Einstein equation admits translationally non-uniform black string solutions. It has been argued that infinitesimally non-uniform black strings should be unstable in 13 or fewer dimensions and otherwise stable. We construct numerically non-uniform black string solutions in 11, 12, 13, 14 and 15 dimensions. Their stability is investigated using local Penrose inequalities. Weakly non-uniform solutions behave as expected. However, in 12 and 13 dimensions, strongly non-uniform solutions appear to be stable and can have greater horizon area than a uniform string of the same mass. In 14 and 15 dimensions all non-uniform black strings appear to be stable. 
\end{abstract}

\section{Introduction}

Consider vacuum General Relativity in $D$ dimensions compactified on a Kaluza-Klein (KK) circle. This theory admits a black string solution given by the product of a $D-1$ dimensional Schwarzschild solution of radius $r_+$ with a circle of circumference $L=2\pi R$.  This black string has horizon topology $S^{D-3} \times S^1$ and is invariant under translations around the $S^1$. Gregory and Laflamme showed that the string is unstable against linearised gravitational perturbations if $r_+<r_*$ for some critical radius $r_*$ proportional to $R$ \cite{Gregory:1993vy}. For example, if $D=5$ then $r_* = 0.8762 R$. 

Some time ago, it was conjectured that this theory also admits {\it non-uniform} black strings (NUBSs), which break the translational invariance around $S^1$, and form a 1-parameter family which bifurcates from the uniform black string family at the marginally stable solution with critical radius $r_*$ \cite{Horowitz:2001cz}. This conjecture has been confirmed by perturbative construction of such solutions, valid for small non-uniformity \cite{Gubser:2001ac,Sorkin:2004qq}, and by numerical solution of the Einstein equation \cite{Wiseman:2002zc,Sorkin:2006wp,Headrick:2009pv}. 

A convenient parameterization of these solutions is with the non-uniformity parameter \cite{Gubser:2001ac}
\be
\lambda = \frac{1}{2} \left( \frac{R_{\rm max}}{R_{\rm min}} - 1 \right)
\ee
 where $R_{\rm max}$ and $R_{\rm min}$ are the maximum and minimum values of the radius of the $S^{D-3}$ of the horizon. It is expected that non-uniform black strings exist for all $\lambda > 0$, and that at $\lambda = \infty$, there is a merger between the family of non-uniform black strings, and a 1-parameter family describing black holes localized on the KK circle, with horizon topology $S^{D-2}$ \cite{Kol:2002xz}. Numerical evidence in support of this expectation has been obtained for $D=5,6$ \cite{Sorkin:2003ka,Kudoh:2004hs,Headrick:2009pv}.

The perturbative construction reveals an interesting change in the behaviour of non-uniform black strings at the critical dimension $D=13$ \cite{Sorkin:2004qq}. For $D \le 13$, infinitesimally non-uniform black strings have a smaller horizon area than a uniform black string of the same mass. This implies that such a solution cannot be the endpoint of the Gregory-Laflamme instability of the uniform string. However, for $D>13$, an infinitesimally non-uniform solution has a greater horizon area than a uniform string of the same mass, and is a natural candidate for the endpoint of the GL instability. 

The aim of this paper is to investigate the stability of non-uniform black strings with finite $\lambda$. The perturbative results just discussed suggest that weakly non-uniform (i.e. small $\lambda$) black strings should be unstable for $D \le 13$ and stable for $D>13$. The stability of strongly non-uniform black strings has not been investigated but it has been suggested that such strings might be stable for $D \lesssim13$ \cite{Horowitz:2011cq}.  

Before we can investigate stability, we must determine the relevant non-uniform black string solutions numerically. We are especially interested in strongly non-uniform strings with $D \sim 13$. Such solutions have not been constructed previously. Ref. \cite{Sorkin:2006wp} gave results for $6\le D \le 11$. We will construct solutions with $D \ge 11$. For $D=11$ our solutions have much greater $\lambda$ than those of Ref.  \cite{Sorkin:2006wp}.

Our solutions exhibit some novel features. For $D \le 13$, all non-uniform strings constructed previously have the property that, as $\lambda$ increases, the mass increases. However, for $D=12,13$ we find that the mass increases to a maximum at $\lambda=\lambda_1$ and then decreases for $\lambda>\lambda_1$.\footnote{Ref. \cite{Kleihaus:2006ee} found an analogous feature in $D=5,6$. However, subsequent work  \cite{Sorkin:2006wp} could not reproduce this feature and neither can we. We believe that the result of Ref. \cite{Kleihaus:2006ee} arises from numerical error.} The first law implies that the horizon area exhibits the same qualitative behaviour as the mass, i.e., it is maximized at $\lambda=\lambda_1$. Furthermore, there exists $\lambda_2 > \lambda_1$ such that a NUBS with $\lambda>\lambda_2$ has greater horizon area than a uniform string of the same mass. Hence such NUBSs are thermodynamically preferred in the microcanonical ensemble.

\begin{figure}[h!]
\begin{center}
\includegraphics[scale=0.5]{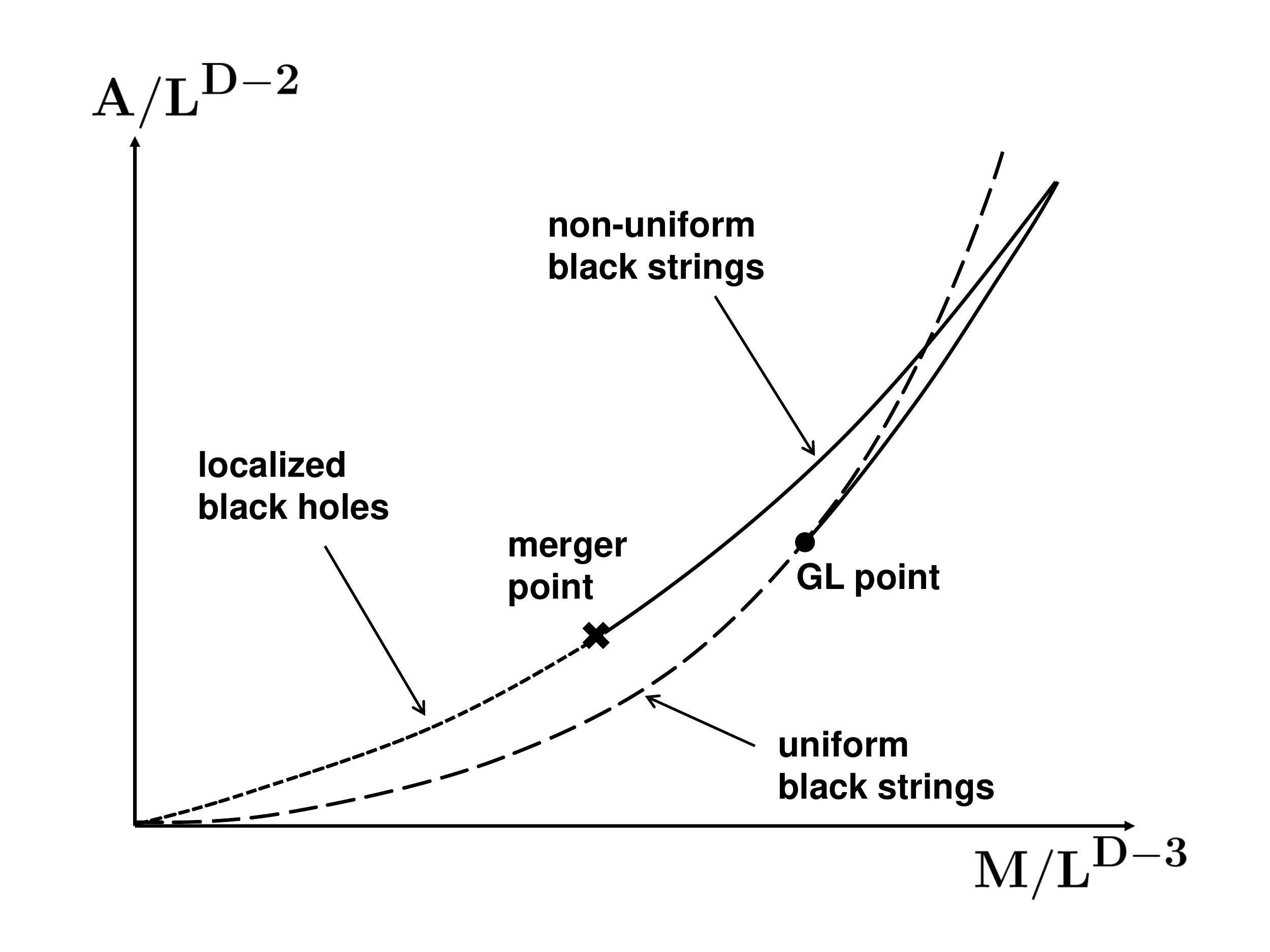}
\end{center}
\caption{Schematic plot of horizon area against mass for black holes and black strings with $D=12,13$. The dashed curve is the uniform black string branch, the solid curve the non-uniform string branch and the dotted curve the localized black hole branch. We find that a maximum mass/area solution occurs on the non-uniform black string branch, resulting in a cusp on the black curve. The localized black hole curve is conjectural but it seems likely that this will merge with the NUBS curve.
}
 \label{figure:phasediagram1}
\end{figure}

\begin{figure}[h!]
\begin{center}
\includegraphics[scale=0.5]{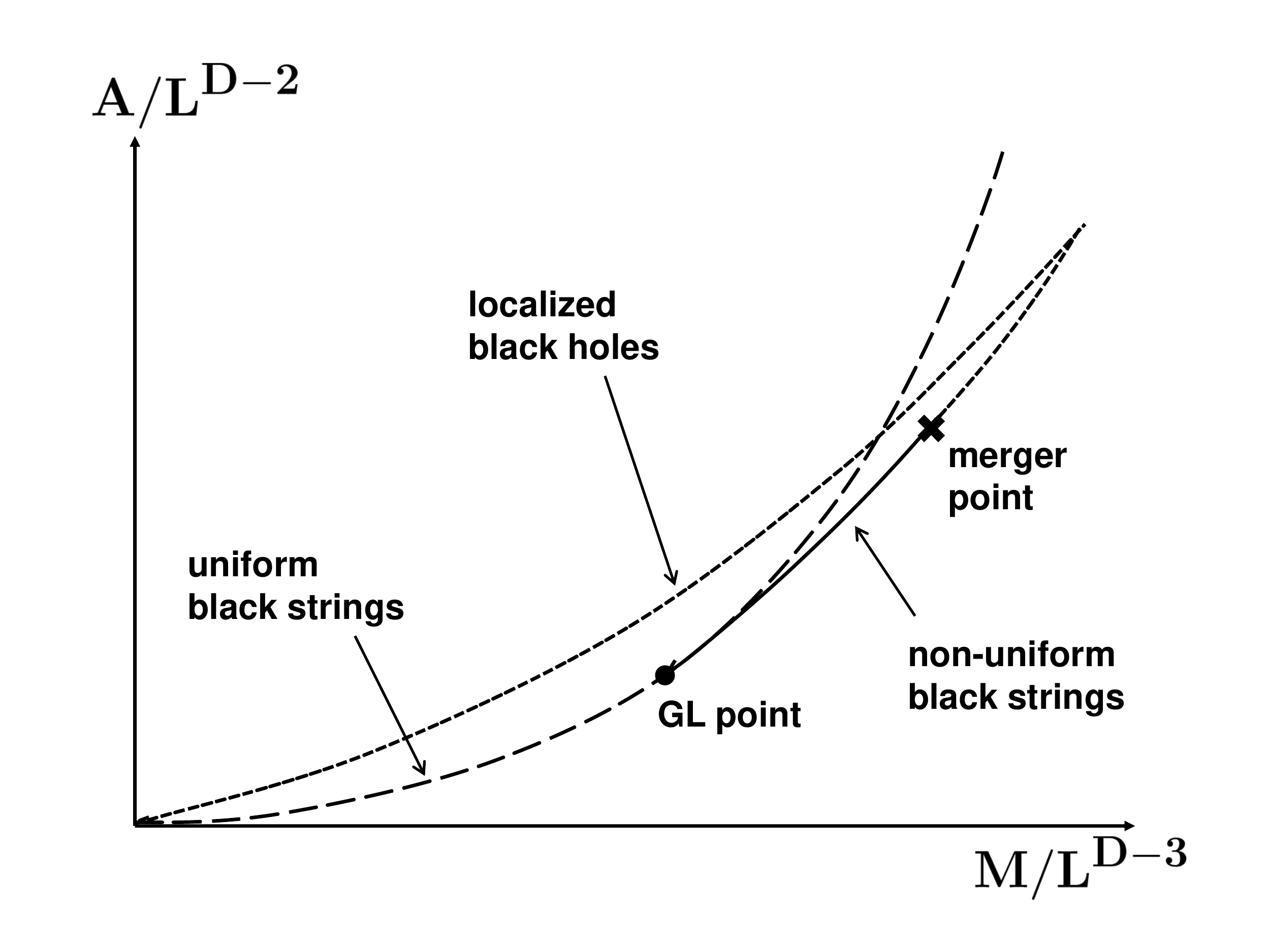}
\end{center}
\caption{Same plot as figure \ref{figure:phasediagram1} but for $D=5,6$, based on results of Refs \cite{Kudoh:2004hs,Headrick:2009pv}. The maximum mass solution (the cusp) occurs on the localized black hole branch. Our results suggest that this will be the qualitative behaviour for all $D \le 11$.
 }
 \label{figure:phasediagram1a}
\end{figure}

\begin{figure}[h!]
\begin{center}
\includegraphics[scale=0.5]{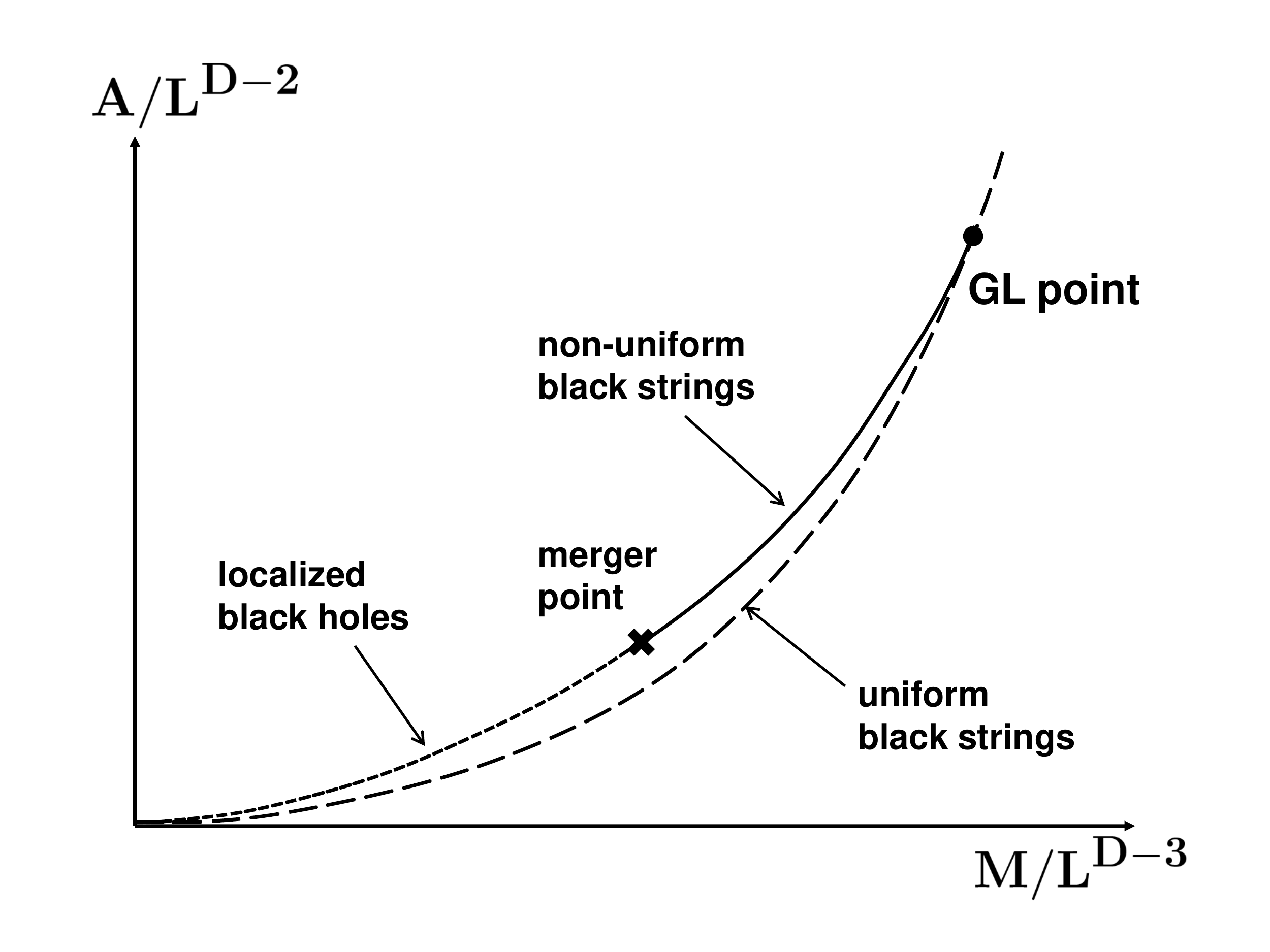}
\end{center}
\caption{Same plot as figure \ref{figure:phasediagram1} but for $D>13$. The behaviour of the localized black hole branch is conjectural.}
\label{figure:phasediagram2}
\end{figure}

Figure \ref{figure:phasediagram1} gives a sketch of these results.\footnote{We have done the plots in Figures \ref{figure:phasediagram1}, \ref{figure:phasediagram1a} and  \ref{figure:phasediagram2} with the real data (for the NUBS and the uniform black strings) but the different phases are so close to each other that they cannot be easily distinguished by eye.} For $D=12,13$, the occurrence of a maximum mass/area along the NUBS branch results in a cusp when horizon area is plotted against mass. We indicate also the expected behaviour of the (not yet constructed) branch of localized black hole solutions. 

The behaviour of the NUBS branch that we have discovered for $D=12,13$ is similar to the behaviour of localized black hole solutions for $D=5,6$. Refs. \cite{Kudoh:2004hs,Headrick:2009pv} found that as one moves along the branch of such solutions, starting from a small black hole, the mass increases to a maximum and then decreases before the merger with the NUBS branch. The resulting phase diagram is shown in Figure \ref{figure:phasediagram1a}, with the cusp appearing on the localized black hole branch. 

It would be very interesting if a turning point (i.e. a maximum of mass/area) exists along the NUBS branch for $D\leq 11$. However, our results for $D=11$ suggest that this is not the case. Our solutions extend all the way up to $\lambda \sim 2.4$ and we have not seen any evidence of a turning point. In fact, when we plot the physical quantities against $\lambda$, we find that for $\lambda\gtrsim 1$ they appear to be nearly constant, which is the expected behaviour as one approaches a merger point. Identifying the value of $D_0$ such that  solutions with $D \le D_0$ have the same qualitative behaviour as for $D=5,6$ (i.e. with a turning point along the localized black hole branch instead of the NUBS branch) will require constructing the localized black hole solutions, but our results show $D_0 < 12$ and suggest that $D_0=11$.

Turning to the case of $D>13$, we have obtained NUBS solutions for $D=14,15$. This is the first time such solutions have been constructed non-perturbatively. We do not discover any surprises with these solutions: as $\lambda$ increases, the mass decreases with no sign of any turning point. This suggests that the phase diagram should take the form of Figure \ref{figure:phasediagram2}, as has been suggested previously \cite{Sorkin:2006wp}.

We will investigate stability using the method of local Penrose inequalities of Ref. \cite{Figueras:2011he}, which we will review below. This method reveals that a necessary condition for stability of a static black hole is that it should be a local maximum of horizon area at fixed mass in the space of initial data for Einstein's equation obeying the relevant asymptotic boundary conditions. Therefore an instability can be demonstrated by constructing initial data describing a perturbed black hole which has greater horizon area than an unperturbed black of the same mass. 

Using local Penrose inequalities, we find that $D \le 11$ NUBSs are unstable for all values of $\lambda$ for which we have constructed solutions. This suggests that all $D \le 11$ NUBSs are classically unstable.

For $D=12,13$, we find that all solutions with $\lambda<\lambda_1$ are unstable. For $\lambda>\lambda_1$, we have constructed multi-parameter families of initial data and found them all to respect the local Penrose inequality, which strongly suggests that such solutions are stable.\footnote{An instability for $\lambda$ close to, but less than, $\lambda_1$ is suggested by the more heuristic "turning point method", which has been used to predict neutron star instabilities \cite{Friedman:1988er}. For black holes, the local Penrose inequality explains the success of this method.} 
Hence for  $D=12,13$, there is a range of masses $M$ for which there exist stable non-uniform black string solutions. In fact, there can exist two stable black string solutions: one uniform and one non-uniform with the non-uniform string entropically preferred for $\lambda>\lambda_2$.

For our $D=14,15$ NUBSs, we have investigated multi-parameter families of initial data and found them all to be consistent with the local Penrose inequality, for all values of $\lambda$ for which we have constructed solutions. This 
strongly suggests that these solutions are stable. 

The organisation of the rest of the paper is as follows. In \S2 we will describe our numerical construction of NUBSs in arbitrary number of dimensions.  Our construction  is based on Ref. \cite{Headrick:2009pv}, which we briefly review in \S\ref{subsec:method} . In \S\ref{subsec:resultsNUBS} we present our results for the different values of $D$. In \S\ref{LPI} we apply the local Penrose inequality method of \cite{Figueras:2011he} to NUBSs. The results on the stability of these solutions are presented in \S\ref{subsec:instability}. 

\section{Numerical construction of NUBSs}
\label{NumericalNUBS}

\subsection{Method}
\label{subsec:method}

In this section we outline our numerical construction of NUBSs in an arbitrary number of space-time dimensions $D$. We will follow the numerical method of \cite{Headrick:2009pv} based on the harmonic formulation of the Einstein equations for finding static black hole space times (see Chapter 10 in \cite{Horowitz:2011cq} for a review). 

Following \cite{Headrick:2009pv} we write the most general ansatz for the metric with the desired isometries ($U(1)_\tau\times SO(D-2)$ in our case):\footnote{Here we work in Euclidean signature but because the spacetimes we are interested in are static we can trivially Wick-rotate the time coordinate and obtain the solution in Lorentzian signature.}
\begin{equation}
ds^2=4\,r_0^2\,\Delta\,y^2\,e^{T}\,d\tau^2+\frac{r_0^2\,e^{S}}{f(y)^{\frac{2}{D-4}}}\,d\Omega_{(D-3)}^2+e^{A}\,dx^2+\frac{4\,r_0^2\,\Delta\,e^{B}}{f(y)^{\frac{2(D-3)}{D-4}}}\,(dy+y\,f(y)\,F\,dx)^2\,,
\label{eqn:ansatz}
\end{equation}
where $f(\xi)=1-\xi^2$ and $\Delta=\frac{1}{(D-4)^2}$ and $r_0$ are  constants. Here the functions $X=\{T,S,A,B,F\}$ depend on the coordinates $(x,y)$ and they are our unknowns. Setting $T=S=A=B=F=0$  in \eqref{eqn:ansatz} we recover the Schwarzschild black string in $D$ spacetime  dimensions. To bring this latter solution to the standard form in Schwarzschild coordinates one performs the following change of coordinates:
\begin{equation}
y^2=1-\frac{r_0^{D-4}}{r^{D-4}}\,.
\end{equation} 
Therefore, the range of the $y$ coordinate is $y\in [0,1)$. The $x$ coordinate will be our compact coordinate and we can take it to lie in the range $x\in[0,1]$ without loss of generality.  In these coordinates $y=0$ is the horizon and $y=1$ is the asymptotic region. Similarly, $x=0$ is the reflection plane and $x=1$ corresponds to the periodic boundary. Therefore, with this particular choice of coordinates we have effectively set the length of the compact circle to be $L=2$. 

With the ansatz \eqref{eqn:ansatz} we solve (numerically) the harmonic Einstein equations \cite{Headrick:2009pv}
\begin{equation}
R_{ab}-\nabla_{(a}\xi_{b)}=0\,,
\end{equation}
subject to suitable boundary conditions. Here $\xi^a=g^{bc}(\Gamma^a_{\phantom a bc}-\bar \Gamma^a_{\phantom a bc}$) is the DeTurck vector, $\Gamma$ is the Levi-Civita connection associated to our metric \eqref{eqn:ansatz}, and  $\bar\Gamma$ is another connection associated to a reference metric $\bar g$ that we are free to prescribe.\footnote{In principle one could be more general and $\bar\Gamma$ need not be the Levi-Civita connection associated to $\bar g$ \cite{Headrick:2009pv}.} In our construction we have chosen $\bar g$ to be the metric of the Schwarzschild black string in $D$ dimensions.

With our choice of gauge, the boundary conditions that we impose are as follows.  At the horizon ($y=0$) smoothness of the metric requires that the functions $X$ should be even in $y$ and therefore we impose a Neumann boundary condition on all of them. At the reflection plane and the periodic boundary ($x=0,1$ respectively), all functions should be even. Finally, at infinity we impose that our solution is asymptotically KK which implies $X=0$.  With these boundary conditions the maximum principle of \cite{Figueras:2011va} implies that $\xi^a$ should vanish and all the solutions to the harmonic Einstein equations should in fact be Einstein (and more precisely Ricci flat in our case). Therefore, we can use $\xi^a$ to monitor the numerical error of our solutions.  

To solve the equations numerically we have used a Chebyshev collocation approximation. For low dimensions and small $\lambda$ a single patch that covers the whole computational domain is enough to give accurate numerical results. However, as $D$ increases it becomes harder to find accurate numerical solutions because the gradients of the functions $X$ become larger and typically higher resolutions are required. To solve this problem we have used 2 conforming meshes or up to 3 non-conforming meshes to cover the domain.\footnote{Ref. \cite{Sorkin:2006wp} followed a similar approach.} This has allowed us to get enough resolution in the regions where it is needed, which is basically where the string pinches in the $\lambda\to\infty$ limit. For the results presented below the numerical error is estimated (monitoring $\xi^a)$ to be less than 1\%. In addition, we have checked that our numerical solutions converge exponentially to the continuum which, given our numerical scheme, strongly suggests that there exist smooth NUBSs. 

The Einstein vacuum equations are scale invariant and we have exploited this freedom to fix the overall scale by setting $L=2$. Then, in any given $D$, the family of NUBSs is parametrised by the inverse temperature $\beta$ and therefore to move along the family we can vary this quantity.\footnote{
Actually we'll have to be more careful for $D=12$ because $\beta$ does not uniquely characterise the solution.} With our ansatz for the metric \eqref{eqn:ansatz}, the inverse temperature of the solution is given  by
\begin{equation}
\beta=\frac{2\pi\,r_0}{n-1}\;,
\end{equation}
and therefore we can move along the branch of NUBS by varying $r_0$.

Finally we note that because in our construction the metric coefficients are even functions of $y$, we can analytically continue the range of the $y$ coordinate to negative values. This corresponds to going across the bifurcation surface and to the other side of the Einstein-Rosen bridge with another asymptotically KK region being at $y\to -1$. As we shall see in \S\ref{LPI} this will be useful in order to study the stability of the NUBSs via the local Penrose inequality.

\subsection{Results}
\label{subsec:resultsNUBS}

In this subsection we report on our results for NUBSs in $D=11,12,13,14,15$ dimensions.\footnote{The results in $D=15$ are qualitatively similar to those in $D=14$ and we will not display them below. Our data extends up to $\lambda\sim 0.8$. }  As we have explained above, for fixed $D$, we keep the length of the asymptotic KK circle fixed and we move along the branch by varying the (inverse) temperature, starting at the GL point, $\beta=\beta_\textrm{GL}$.

\subsubsection{Non-uniformity}

\begin{figure}[t!]
\begin{center}
\includegraphics[scale=0.65]{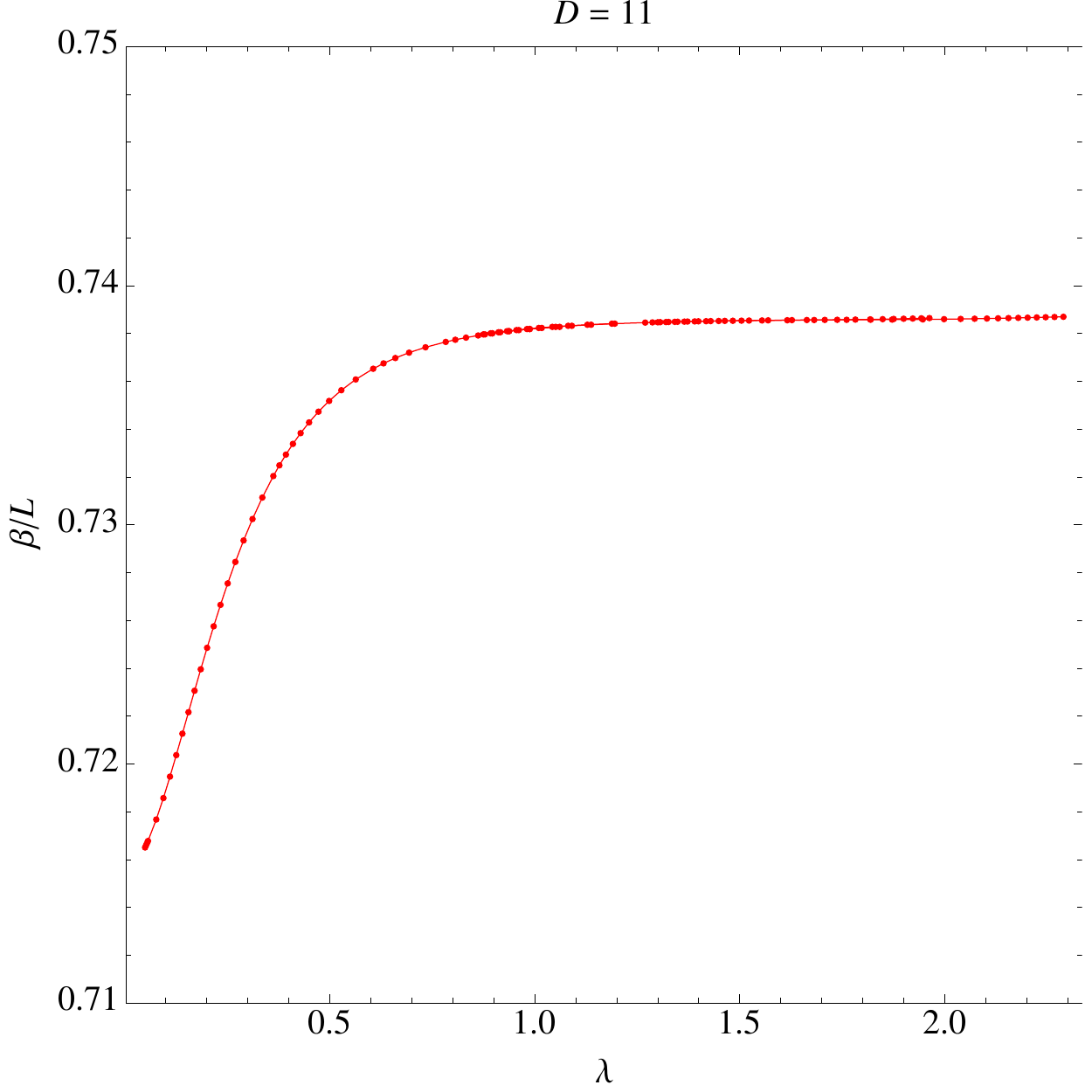}
\includegraphics[scale=0.65]{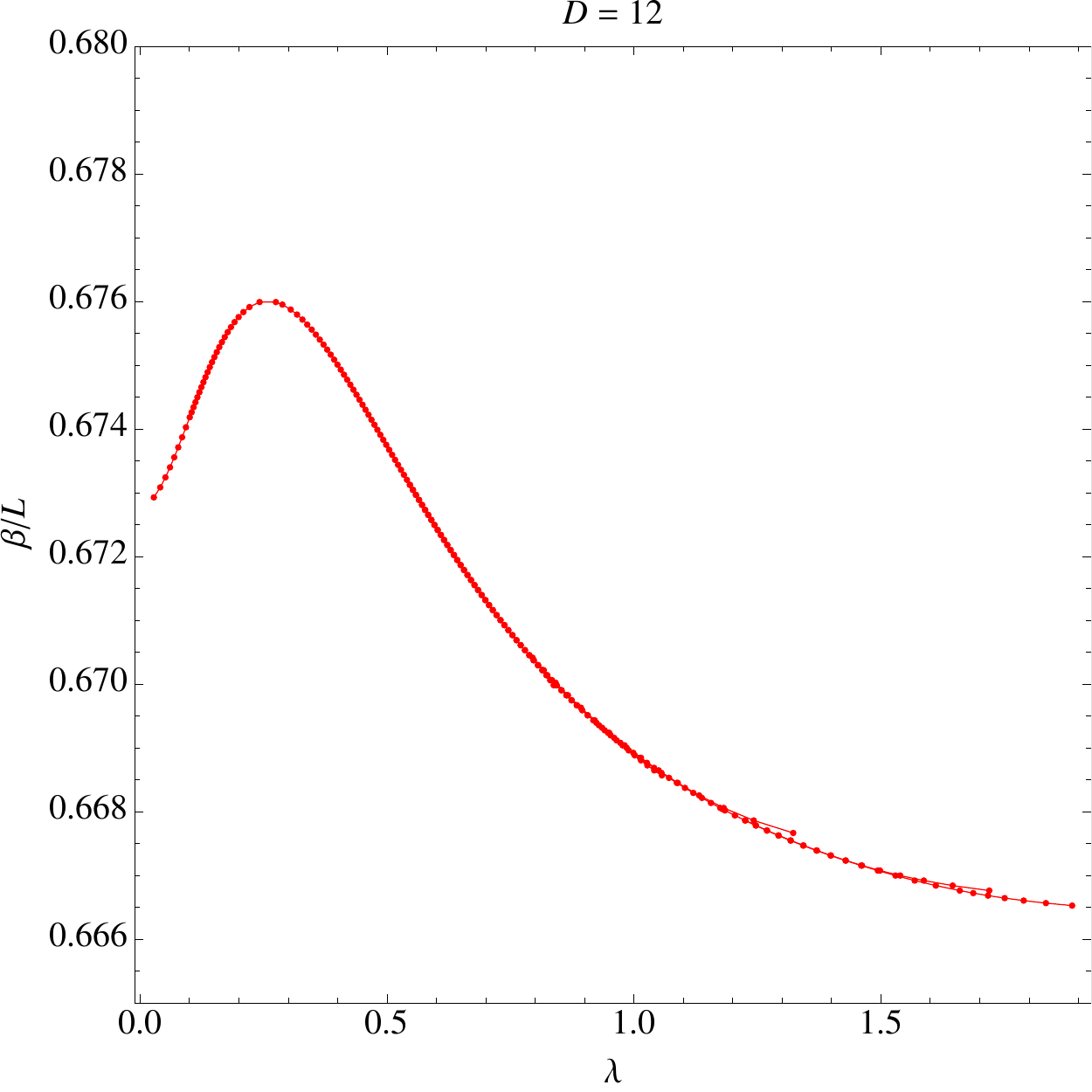}
\\
\includegraphics[scale=0.65]{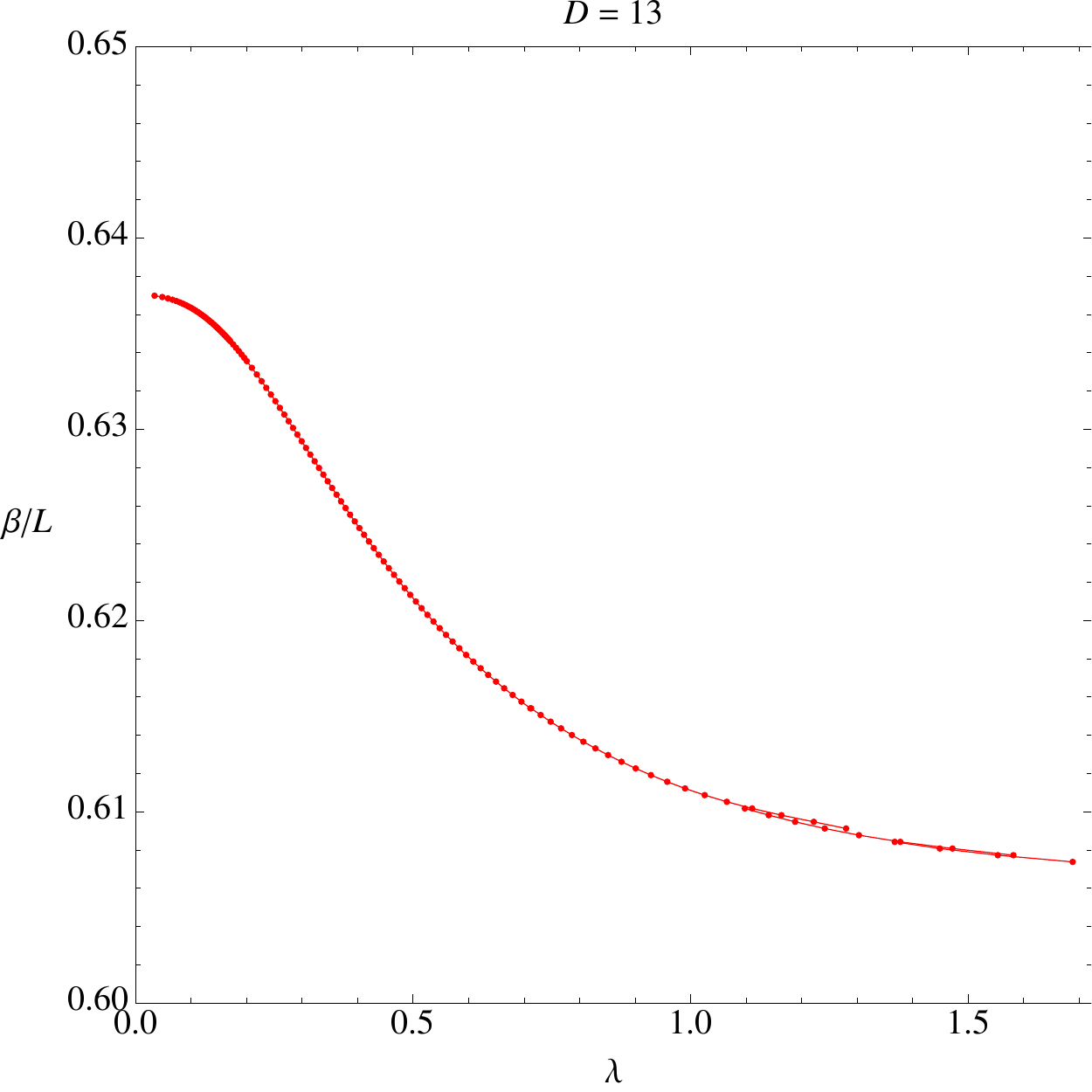}
\includegraphics[scale=0.65]{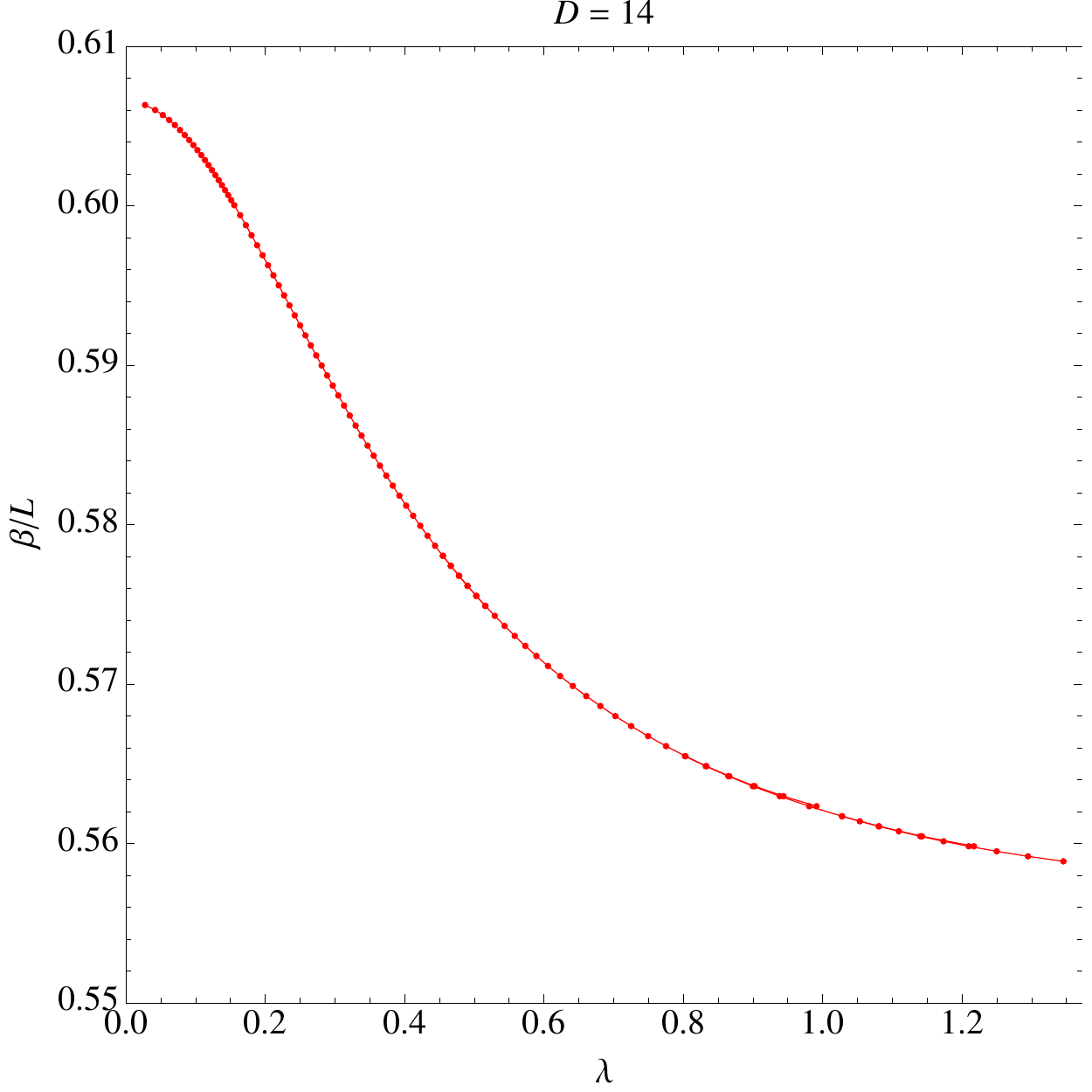}
\end{center}
\caption{$\beta/L$ vs. $\lambda$ for $D=11,12,13,14$. The overlapping curves correspond to data generated at different resolutions. In $D=11$, $\beta$ is a monotonically increasing function of $\lambda$. This is the same behaviour as in the lower dimensional cases \cite{Wiseman:2002zc,Sorkin:2006wp,Headrick:2009pv}.  In $D=12$, $\beta/L$ reaches a maximum at $\lambda=\lambda_0=0.258$ and then it decreases, presumably up to the merger point. In $D\geq 13$, $\beta/L$ is a monotonically decreasing function of $\lambda$.}
\label{fig:betavslambda}
\end{figure}

As just mentioned, we use $\beta$ to parameterize our NUBS solutions when constructing them numerically. However, it is more usual to use $\lambda$ as the parameter and so we start in Figure  \ref{fig:betavslambda}  by showing the dependence of $\beta$ on $\lambda$ for $D=11,12,13,14$. These plots reveal the range of $\lambda$ for which we have constructed solutions. 

For $D=11$, $\beta$ is a monotonically increasing function of $\lambda$, i.e., the temperature decreases with $\lambda$. This is the same as happens in lower dimensions \cite{Wiseman:2002zc,Sorkin:2006wp,Headrick:2009pv}.
The plot flattens out beyond $\lambda \sim 1$, suggesting that $\beta$ is already close to the limiting value it attains at the merger point $\lambda=\infty$. 

For $D=12$ we find qualitatively new behaviour. As we move along the branch of NUBSs starting from the GL point and increasing $\beta$,  we cannot find any solutions beyond $\beta=\beta_0 \equiv 0.6760\,L$. For this solution $\lambda=\lambda_0=0.258$ and therefore it cannot be the end of the branch (which we expect to have $\lambda=\infty$). Instead it is a maximum of  $\beta$, i.e., a minimum of the temperature. Ref. \cite{Headrick:2009pv} found a maximum $\beta$ on the localised black branch for $D=5$; our results show that this maximum of $\beta$ switches to the NUBSs in $D=12$.\footnote{Here we assume that in the yet-to-be constructed localised black hole branch the area of the horizon behaves monotonically as a function of $\beta$. } To find solutions with $\lambda>\lambda_0$ we have used as initial guess for the Newton algorithm a NUBS with $D=13$  and a similar value of $\lambda$.\footnote{For instance, a NUBS in $D=13$ with $\lambda\sim 0.3$ gives a good enough initial guess to find NUBSs in $D=12$ with $\lambda > \lambda_0$.} 

For $D=13,14,15$ $\beta$ is a monotonically decreasing function of $\lambda$, i.e., the temperature increases with $\lambda$.

\subsubsection{Horizon area and mass}
 
\begin{figure}[t!]
\begin{center}
\includegraphics[scale=0.65]{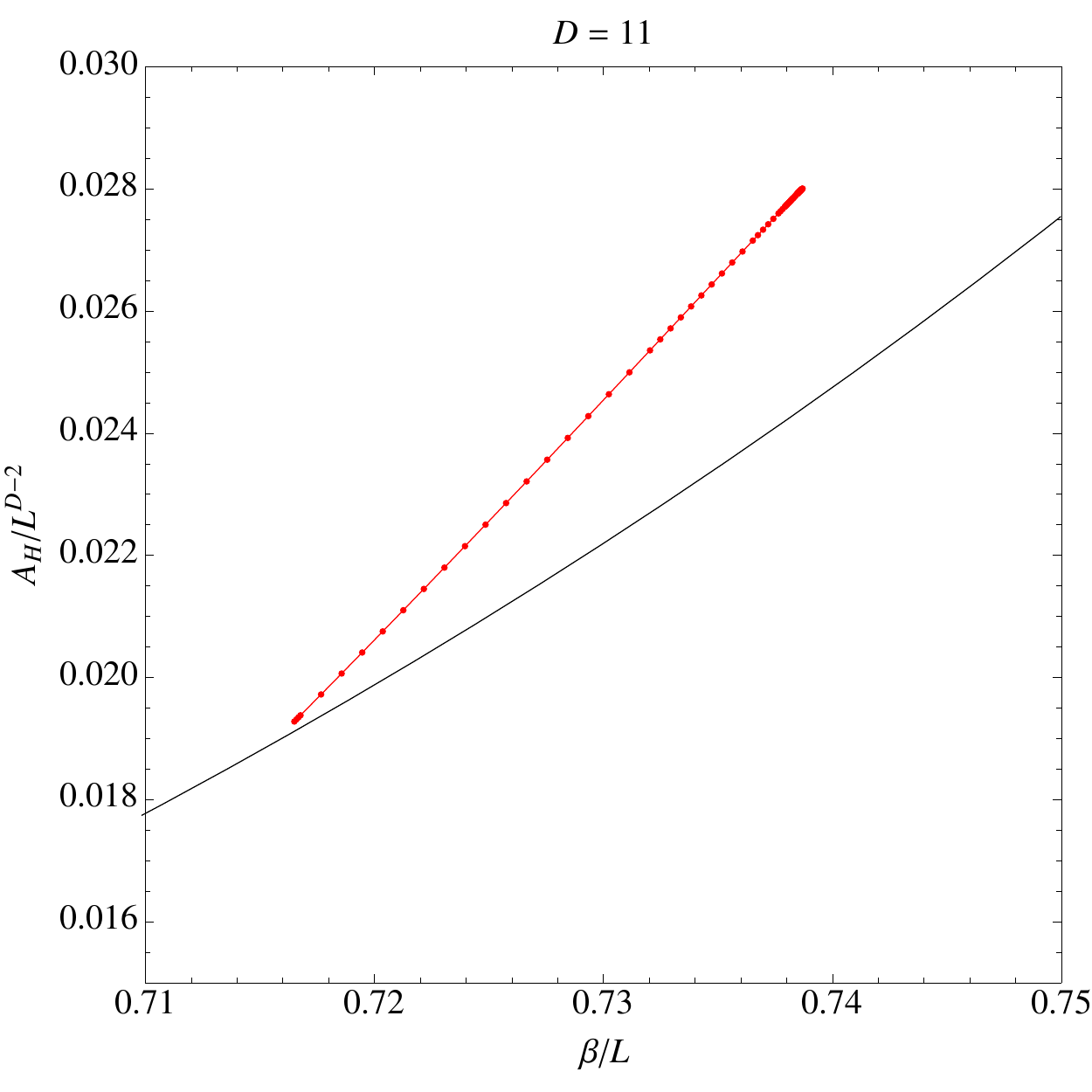}
\includegraphics[scale=0.65]{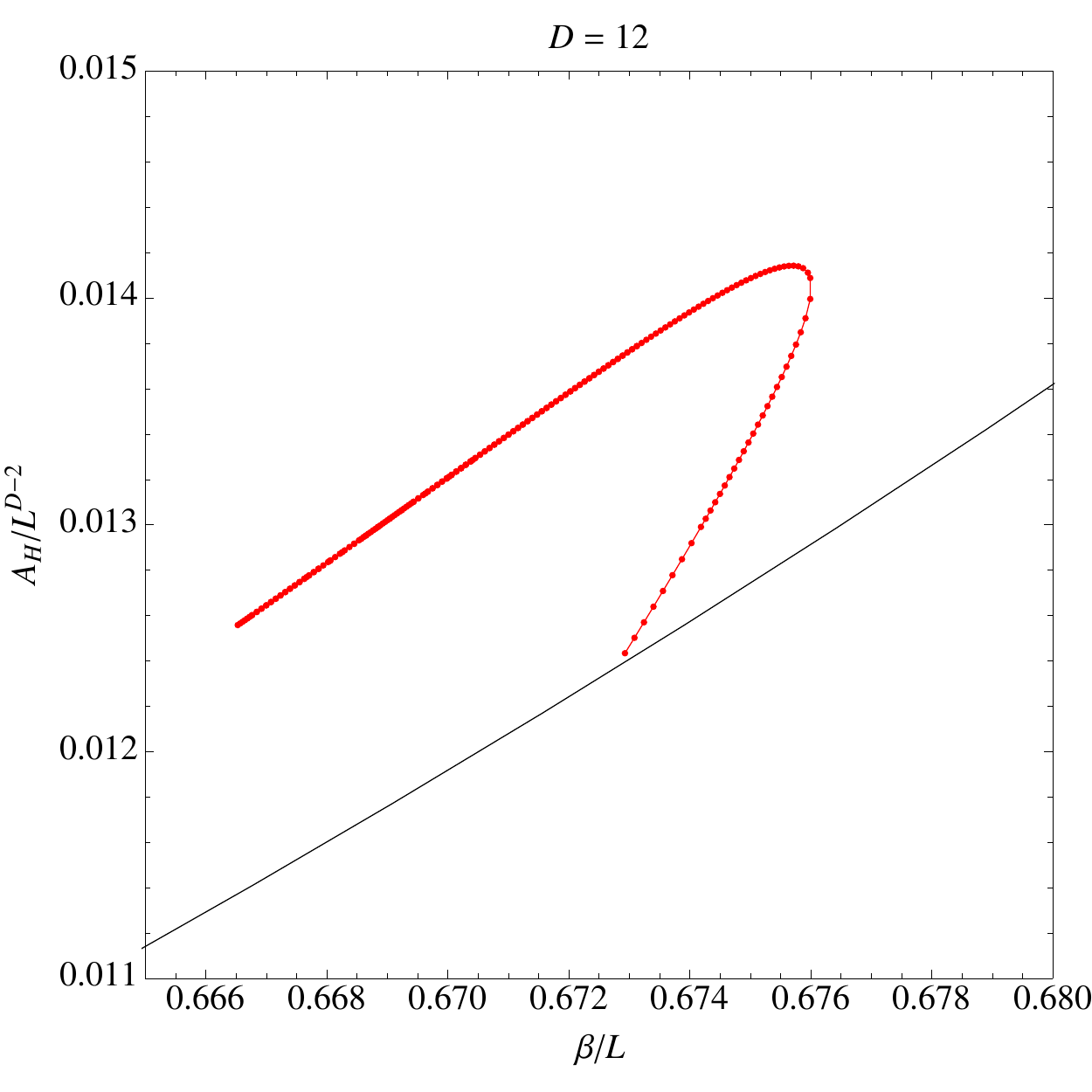}
\\
\includegraphics[scale=0.65]{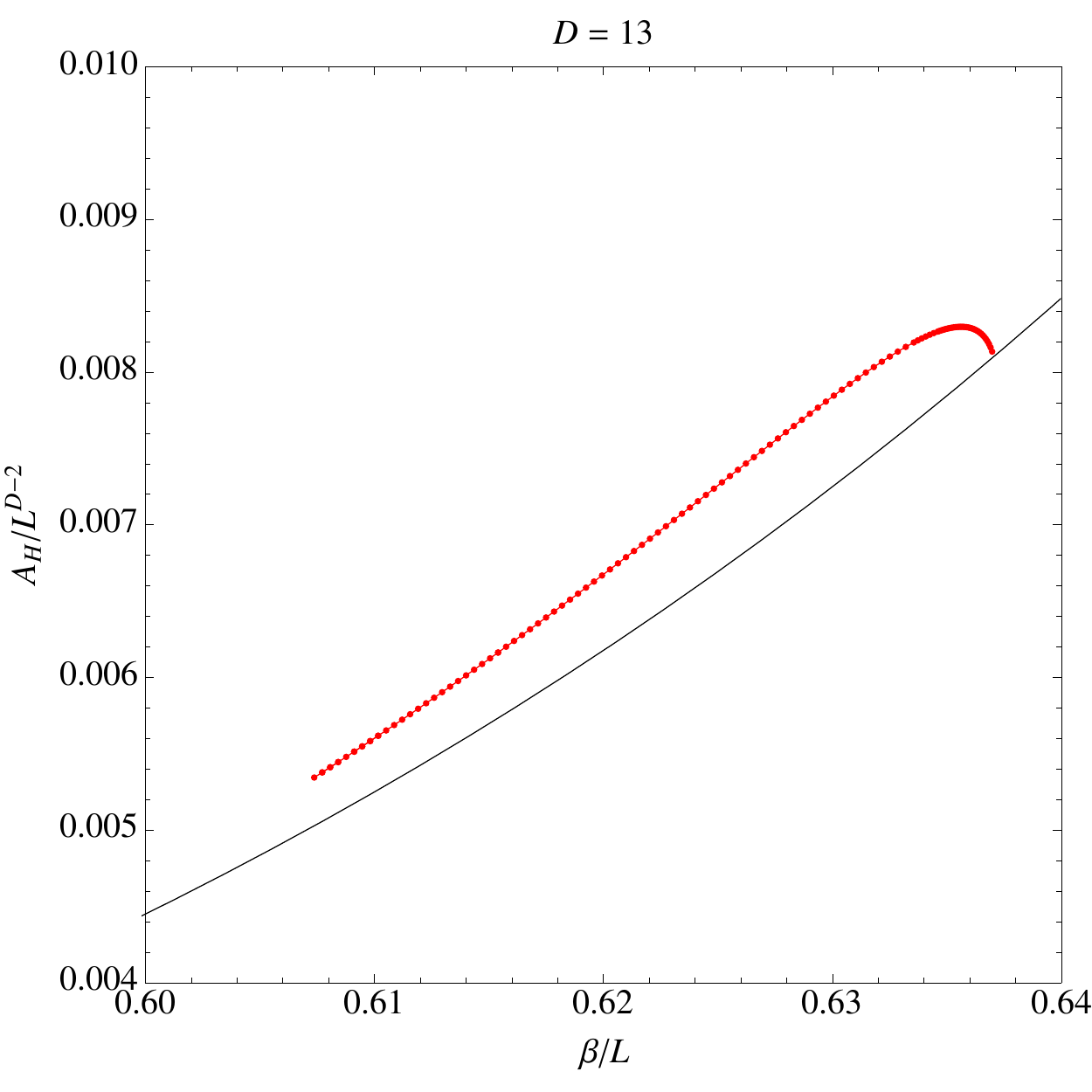}
\includegraphics[scale=0.65]{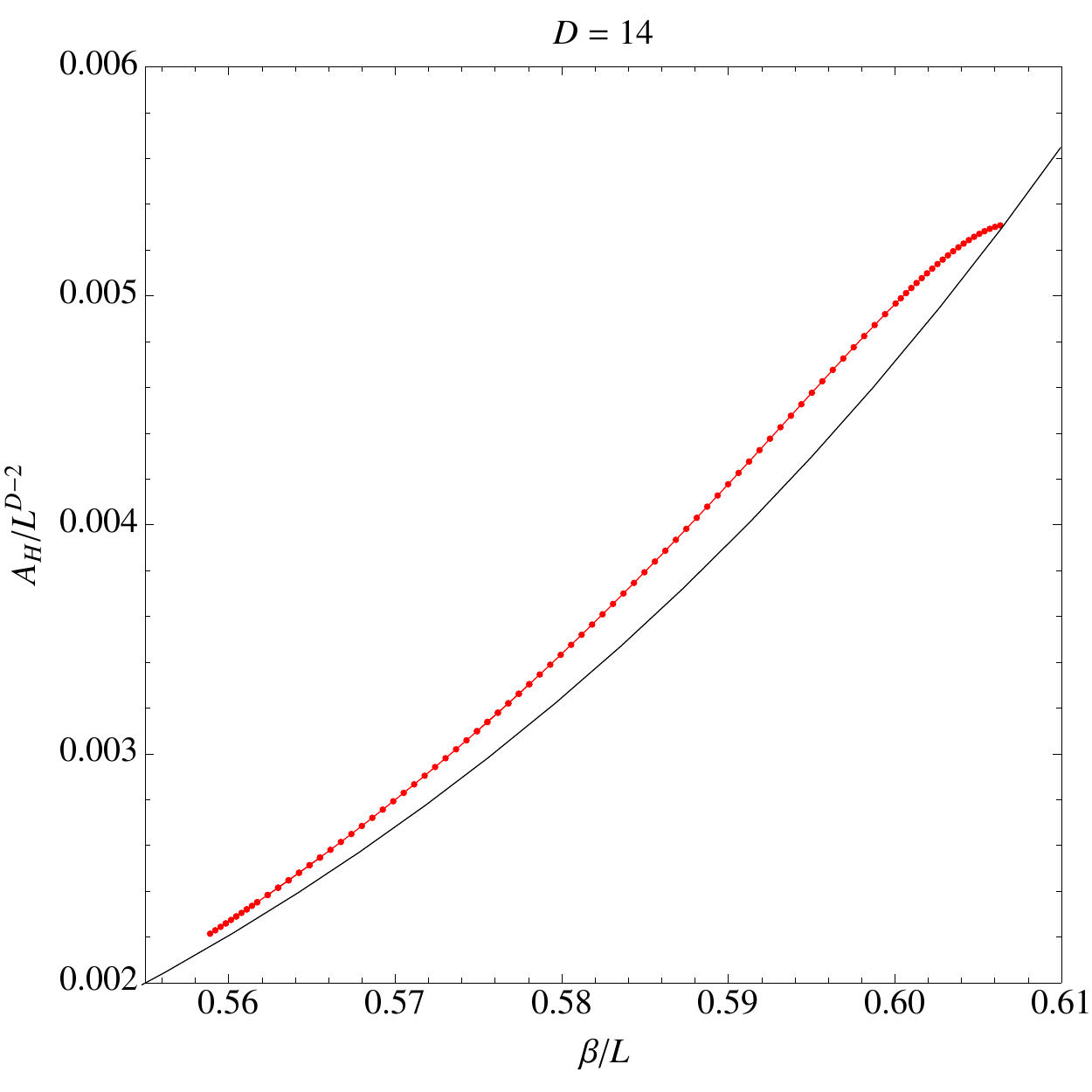}
\end{center}
\caption{$A_H/L^{D-2}$ vs. $\beta/L$ for $D=11,12,13,14$. The black curves are the uniform strings, the red curves are our results for NUBSs. For $D=11$ the area of the horizon is an increasing function of $\beta$, just as in lower dimensional cases previously considered \cite{Wiseman:2002zc,Sorkin:2006wp,Headrick:2009pv}. For $D=12$ there exists a maximum inverse temperature at $\beta_0/L\simeq 0.6760$ and a maximum of the area at $\beta_1/L\simeq 0.6757$.  In $D=13$ the maximum of the area is at $\beta_1/L\simeq 0.6356$ and in $D\geq 14$ the area always decreases as $\beta$ decreases. By the First Law the mass has the same qualitative behaviour as the horizon area. }
\label{fig:resultsarea}
\end{figure}

\begin{figure}[t!]
\begin{center}
\includegraphics[scale=0.65]{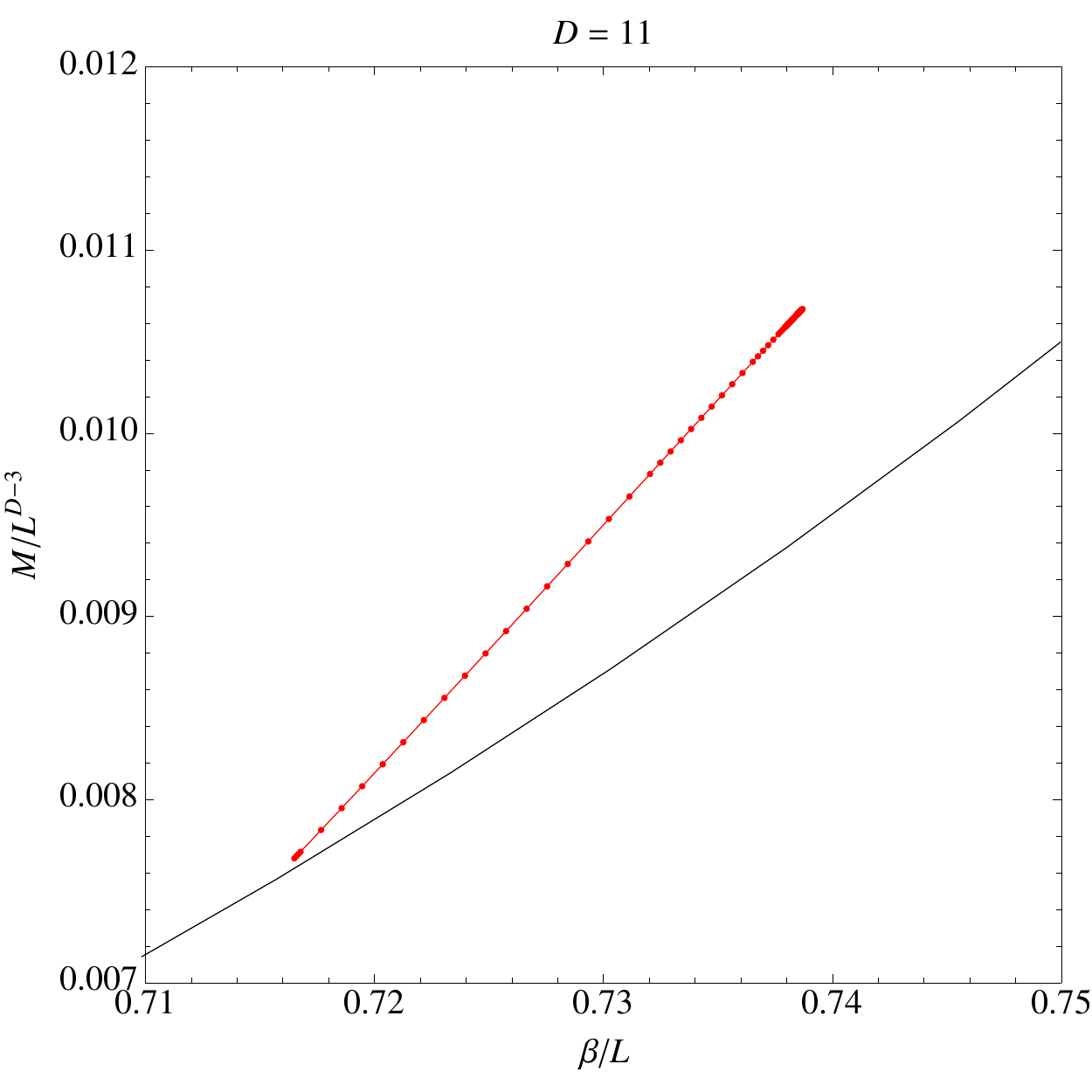}
\includegraphics[scale=0.65]{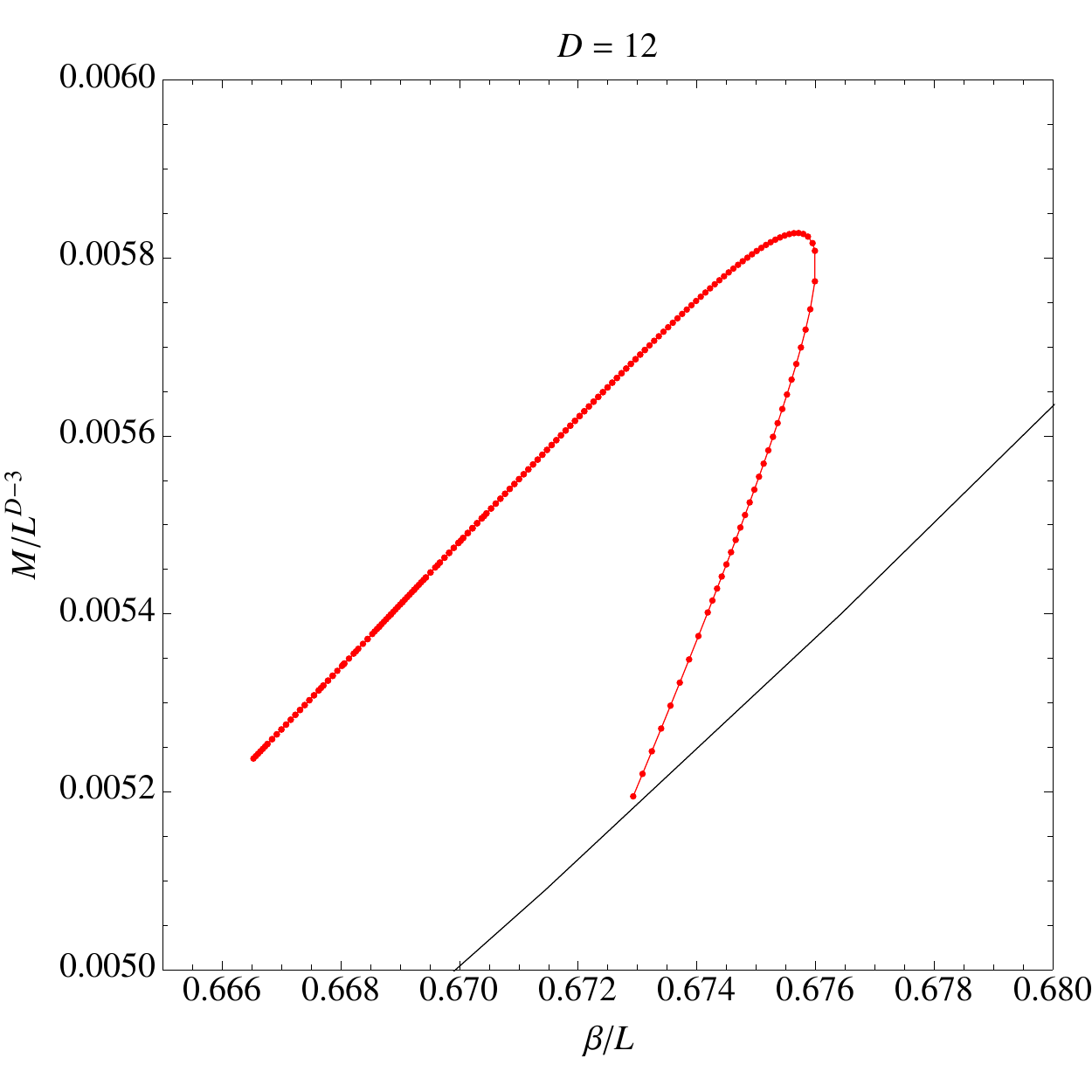}
\\
\includegraphics[scale=0.65]{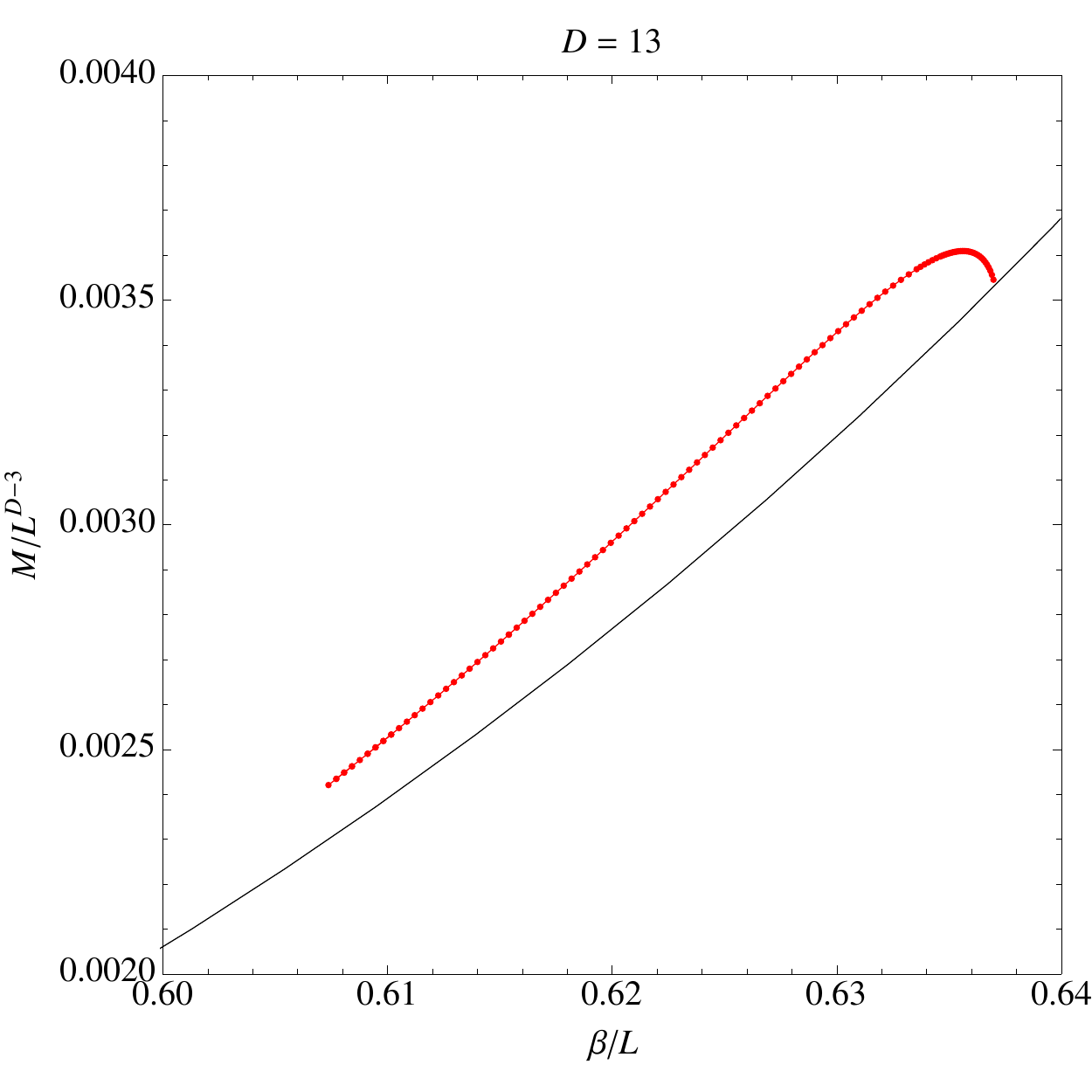}
\includegraphics[scale=0.65]{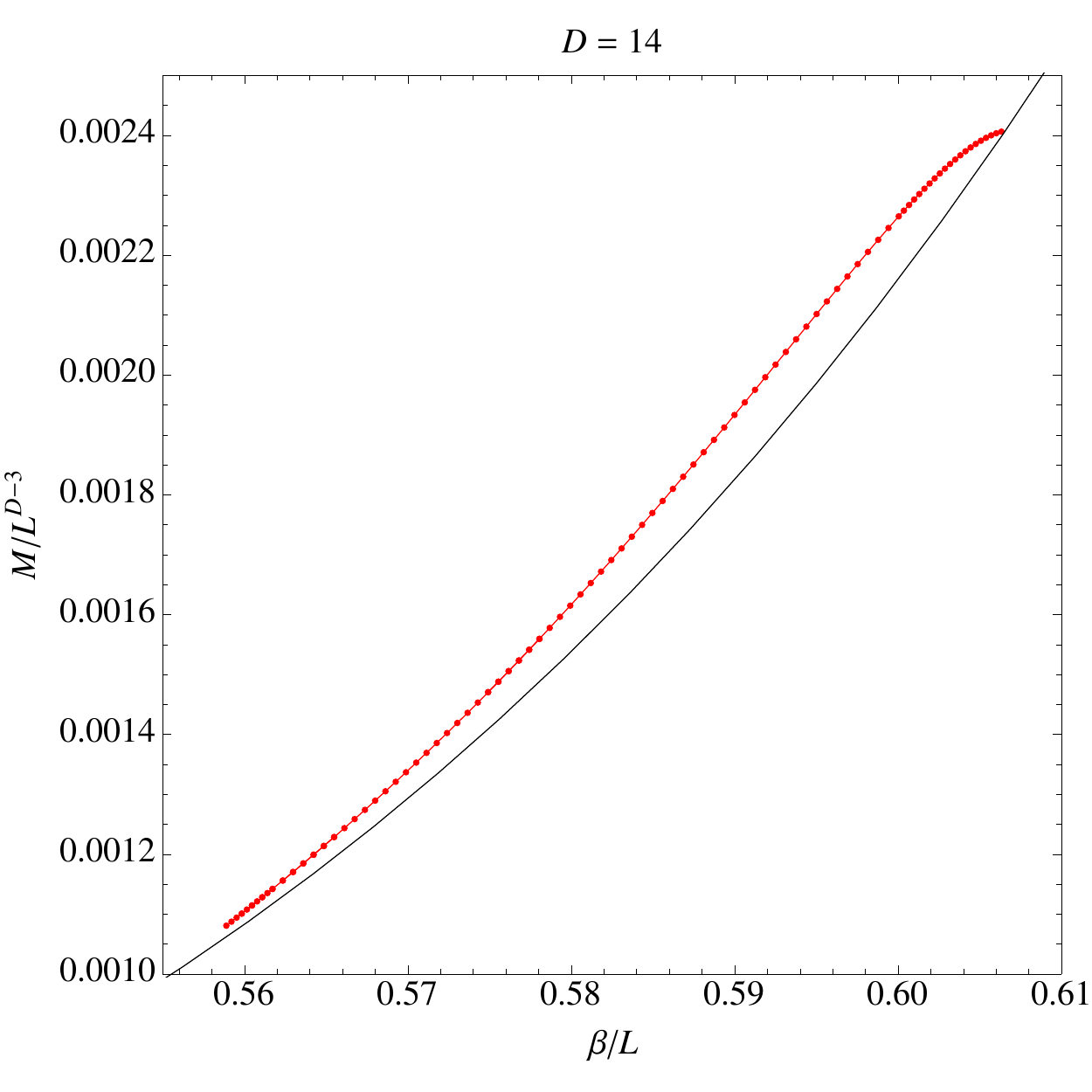}
\end{center}
\caption{$M/L^{D-3}$ vs. $\beta/L$ for $D=11,12,13,14$. The black curves are the uniform black strings. The behaviour of $M$ as a function of $\beta$ in various dimensions is similar to that of horizon area $A_H$, as dictated by the first law. }
\label{fig:resultsmass}
\end{figure}

In Figure \ref{fig:resultsarea}  we plot $A_H/L^{D-2}$ vs. $\beta/L$ for various dimensions ($A_H$ is the horizon area) and for completeness in Figure \ref{fig:resultsmass} we plot $M/L^{D-3}$ vs. $\beta/L$. The first law implies that the $A_H$ and $M$ exhibit the same qualitative behaviour. Using these data, we have checked that the error in the Smarr relation is consistent with our estimates based on the norm of the  DeTurck vector $\xi^a$. 

For $D=11$,  $A_H$ is a monotonically increasing function of $\beta$ (or $\lambda$), at least up to $\lambda \sim 2.4$ but it is reasonable to expect that this is true for all values $\lambda$ up to the merger point. This qualitative behaviour is the same as in the lower dimensional cases discussed previously \cite{Wiseman:2002zc,Sorkin:2006wp,Headrick:2009pv}. 

For $D=12$ we find qualitatively new behaviour. As we move along the branch of NUBSs starting from the GL point and increasing $\beta$,  the area of the horizon increases up to, and beyond, the minimum temperature (maximum $\beta$) solution. However, it reaches a maximum at $\beta=\beta_1\equiv 0.6757\,L$. The value of $\lambda$ at this point is $\lambda=\lambda_1 = 0.332$. The horizon area decreases for larger $\lambda$. By the first law, the mass also has a maximum at this value of $\beta$ (or $\lambda$). As in the localised black hole case for $D=5$ \cite{Headrick:2009pv}, $\beta_1$ is close to $\beta_0$, but they do not coincide: $\beta_1/\beta_0 \simeq 0.9995\pm 0.0001$. 

For $D=13$ we find that there is again a maximum area/mass solution, which has $\beta=\beta_1 \equiv 0.6356\,L$ and $\lambda=\lambda_1 = 0.135$. 

For $D= 14,15$, the area/mass decreases monotonically with $\lambda$, presumably all the way to the merger point. This was the expected behaviour and our results confirm the expectations well in the non-perturbative regime.  However, as $D$ increases it becomes harder to find accurate numerical solutions (within reasonable time and with reasonable computational resources) and we have only been able to explore the branch of NUBSs up  to $\lambda\sim 1$.

\begin{figure}[t!]
\begin{center}
\includegraphics[scale=0.65]{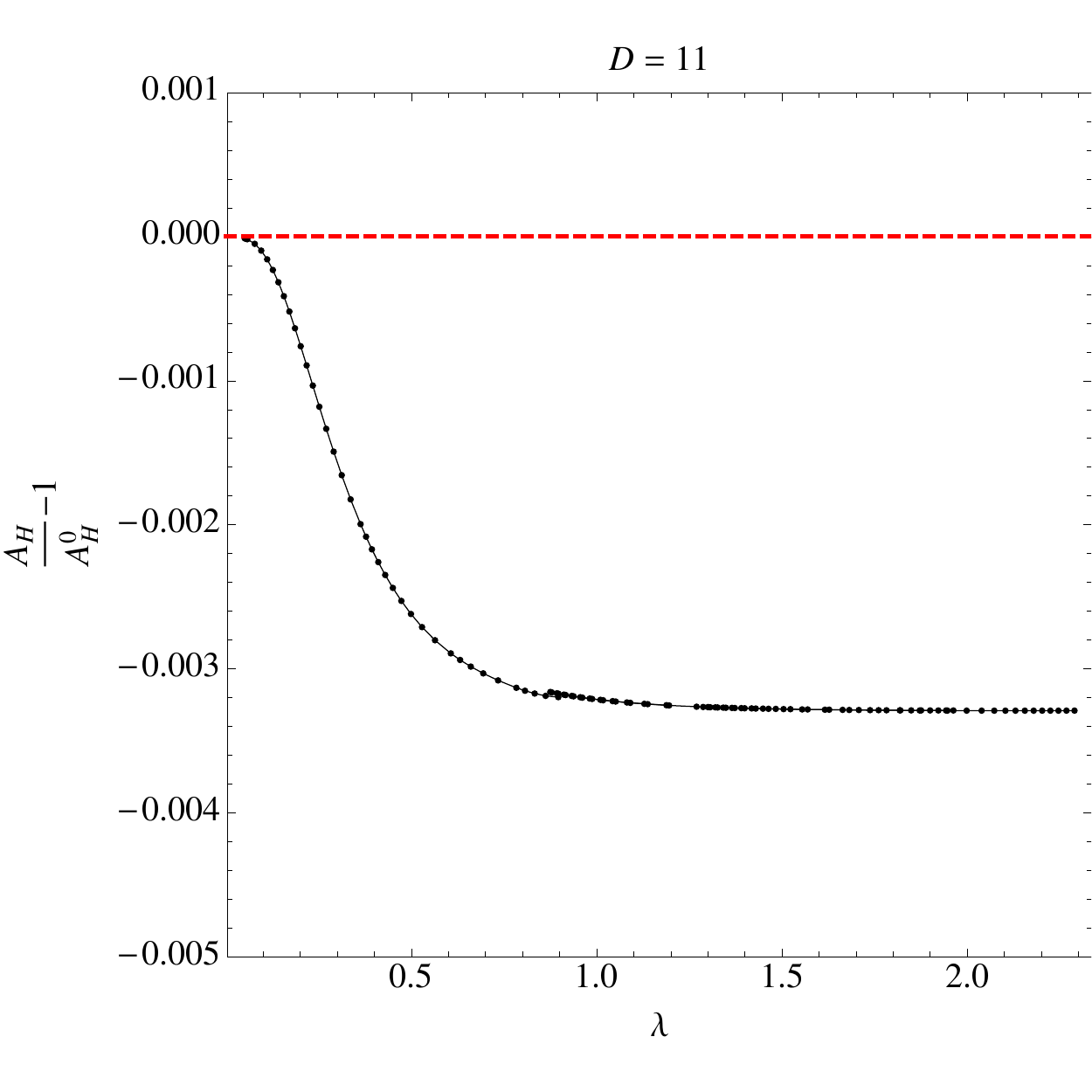}
\includegraphics[scale=0.65]{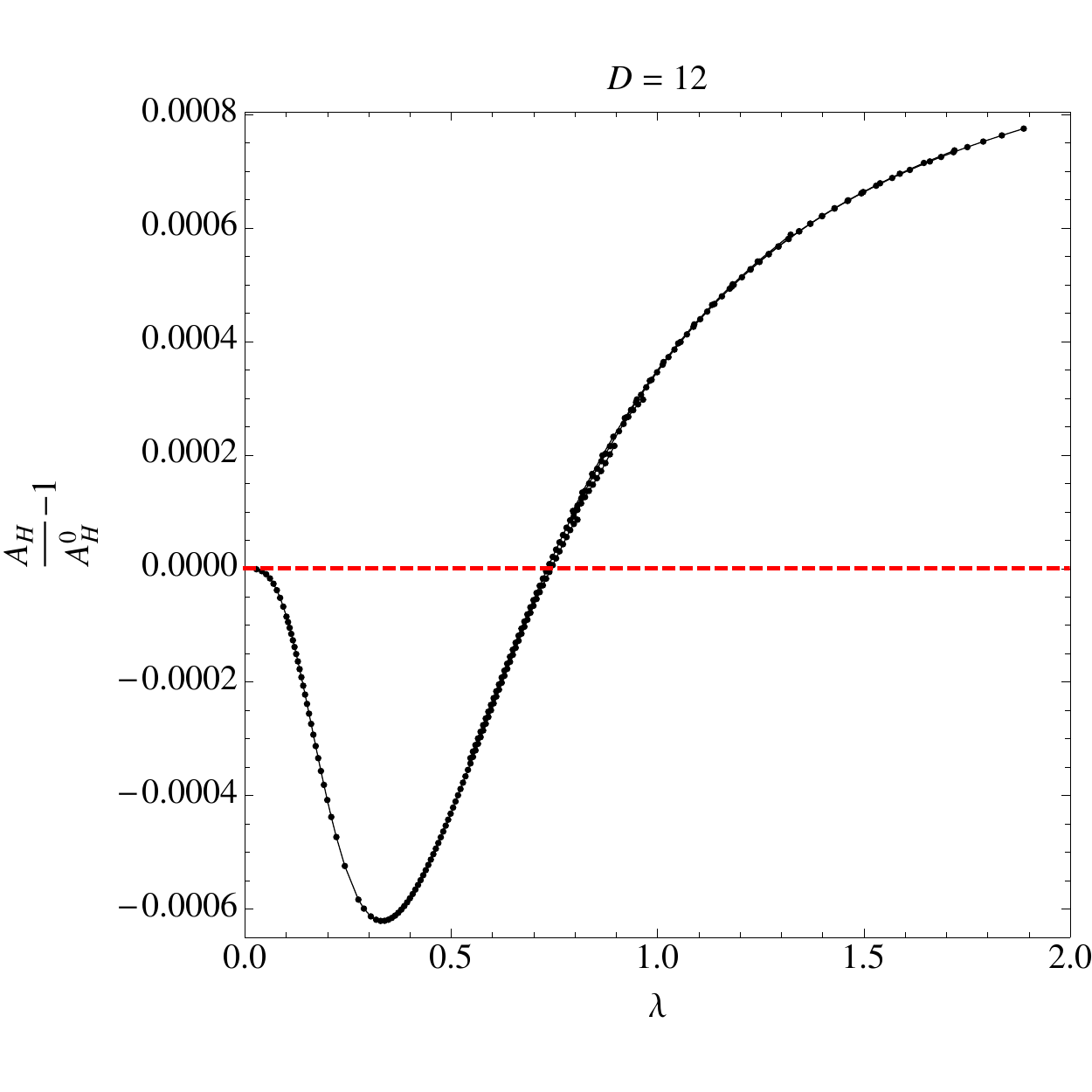}
\\
\includegraphics[scale=0.65]{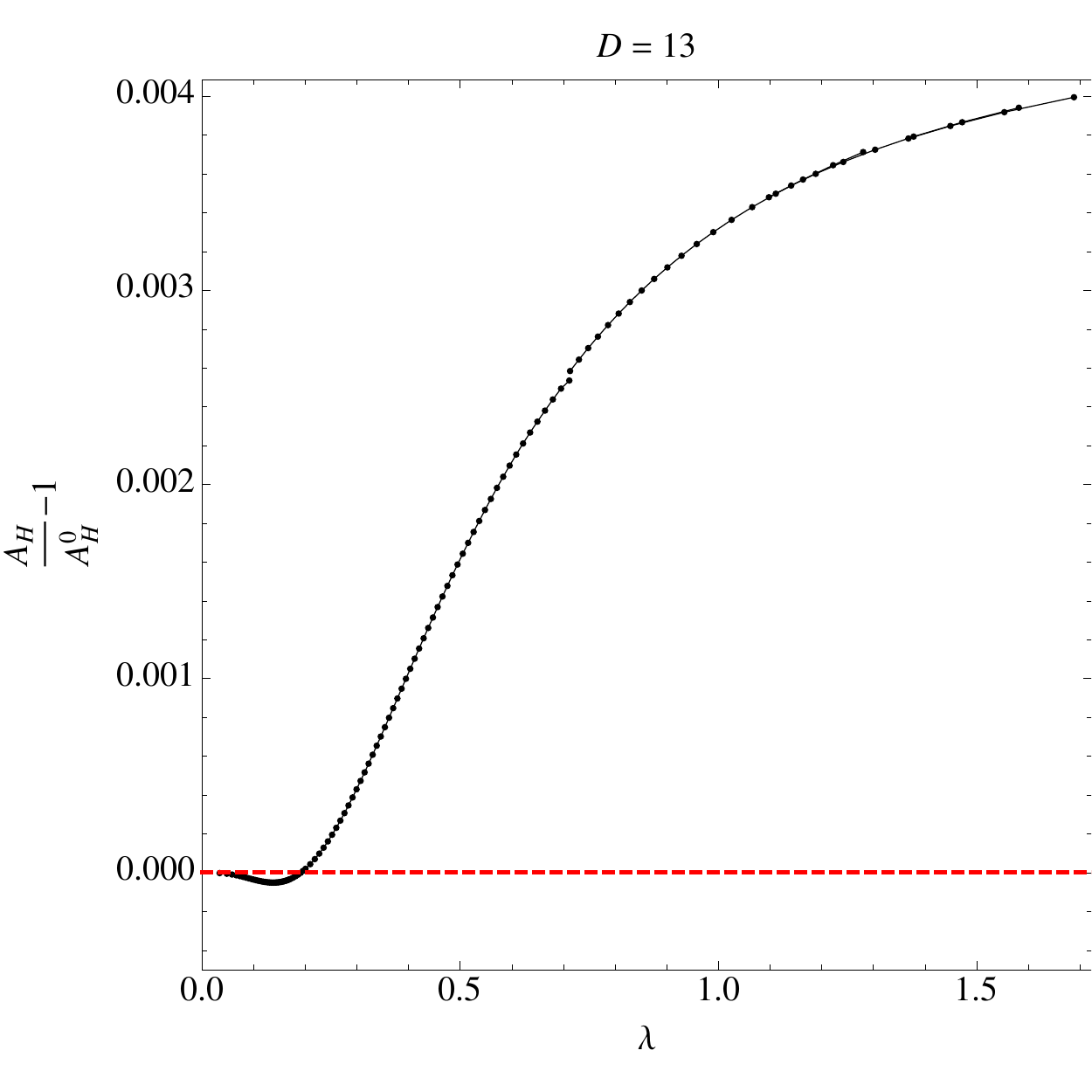}
\includegraphics[scale=0.65]{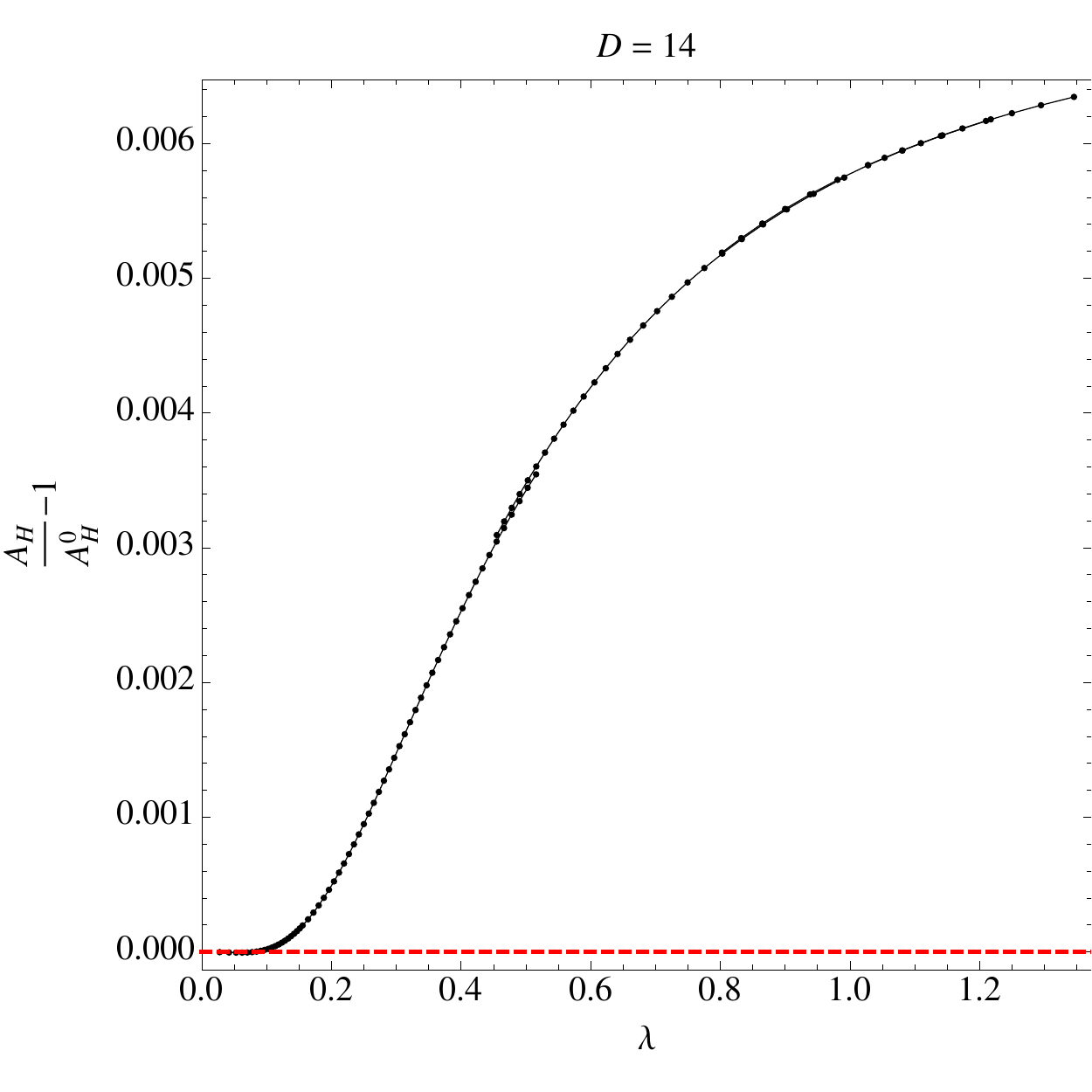}
\end{center}
\caption{Comparing the horizon area of a NUBS with that of a uniform string of the same mass. The vertical axis is $A_H/A_H^0-1$ where $A_H$ is the NUBS horizon area and $A_H^0$ the horizon area of a uniform string with the same mass.  The overlapping curves in this figure correspond to data at different resolutions.  In $D=11$ we observe the same behaviour as in lower dimensions: the NUBSs have lower entropy than the uniform black strings with the same mass. In $D=12,13$ there exists a $\lambda=\lambda_2$ such that for $\lambda>\lambda_2$ the NUBSs have greater entropy than uniform black strings with the same mass. For $D\geq 14$ and any $\lambda>0$ the NUBSs always dominate the microcanonical ensemble. }
\label{fig:entropyvslambda}
\end{figure}

\subsubsection{Comparison with uniform string}

In Figure \ref{fig:entropyvslambda} we plot the difference in horizon area between NUBSs and uniform black strings with the same mass for various space-time dimensions. For $D=11$ we find that the NUBSs always have lower horizon area than uniform black strings with the same mass, as in the previously studied lower dimensional cases \cite{Wiseman:2002zc,Sorkin:2006wp,Headrick:2009pv}. This implies that a $D\leq 11$  NUBS cannot be the endpoint of the GL instability, as   explicitly demonstrated for $D=5$ in Ref. \cite{Lehner:2010pn}.  

For $D=12,13$ we find that there exists $\lambda_2>\lambda_1$ such that for any $\lambda>\lambda_2$, the NUBSs have greater entropy than the uniform black strings with the same mass. We find $\lambda_2 = 0.733$ in $D=12$ and $\lambda_2 =0.192 $ in $D=13$. Therefore, it is natural to expect that NUBSs with $\lambda > \lambda_2$ could be the endpoint of the GL instability, at least in some region of the phase diagram. 

For $D=14,15$, we find that any NUBS has greater entropy than a uniform black string with the same mass, which agrees with the perturbative results of Ref. \cite{Sorkin:2004qq} and extends them well into the non-perturbative regime. 

\subsubsection{Heat capacity}

Let $c_L = dM/dT$ be the heat capacity, with the subscript $L$ referring to the fact that we are keeping the asymptotic length of the KK circle fixed.

For $D=11$, the horizon area increases with $\beta$ which implies that $c_L$ is negative. This is the same as found previously for $D \le 11$ dimensions \cite{Wiseman:2002zc,Sorkin:2006wp,Headrick:2009pv}.

For $D=12$, writing $c_L=(dM/d\lambda)/(dT/d\lambda)$ shows that $c_L$ changes sign by diverging at the minimum temperature solution $\lambda=\lambda_0$ and changes sign by vanishing at the maximum mass solution $\lambda=\lambda_1$. Hence $c_L$ is negative for $\lambda<\lambda_0$, diverges at $\lambda=\lambda_0$, is positive in the small range $\lambda_0 < \lambda < \lambda_1$, vanishes at $\lambda=\lambda_1$, and is negative for $\lambda>\lambda_1$, presumably all the way to the merger point and also along the localised black hole branch. Ref. \cite{Headrick:2009pv} found that for $D=5$ there is a portion of the localised black hole branch which has positive heat capacity. Our results suggest that solutions with positive heat capacity remains on the localised black hole side up to $D=11$. However, for $D=12$ they switch to the NUBS side.  

For $D=13$, $c_L$ is positive for $\lambda<\lambda_1$, vanishes at $\lambda=\lambda_1$, and is negative for $\lambda>\lambda_1$. Presumably the localised BHs have negative $c_L$. 

For $D=14,15$, all NUBSs that we have been able to construct have negative heat capacity. Presumably this is true also of the localised BHs.

\subsubsection{Free energy}

\begin{figure}[t!]
\begin{center}
\includegraphics[scale=0.65]{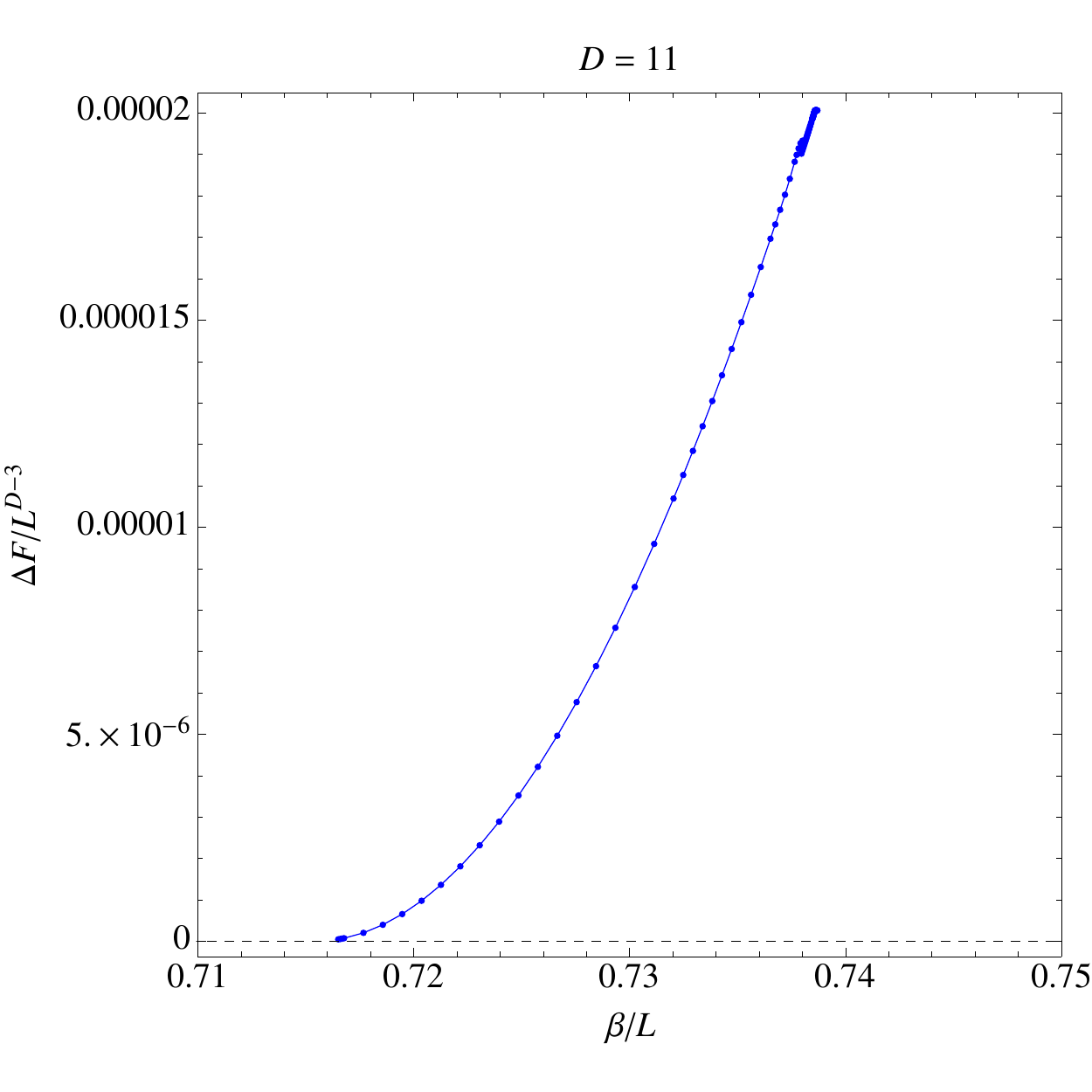}
\includegraphics[scale=0.65]{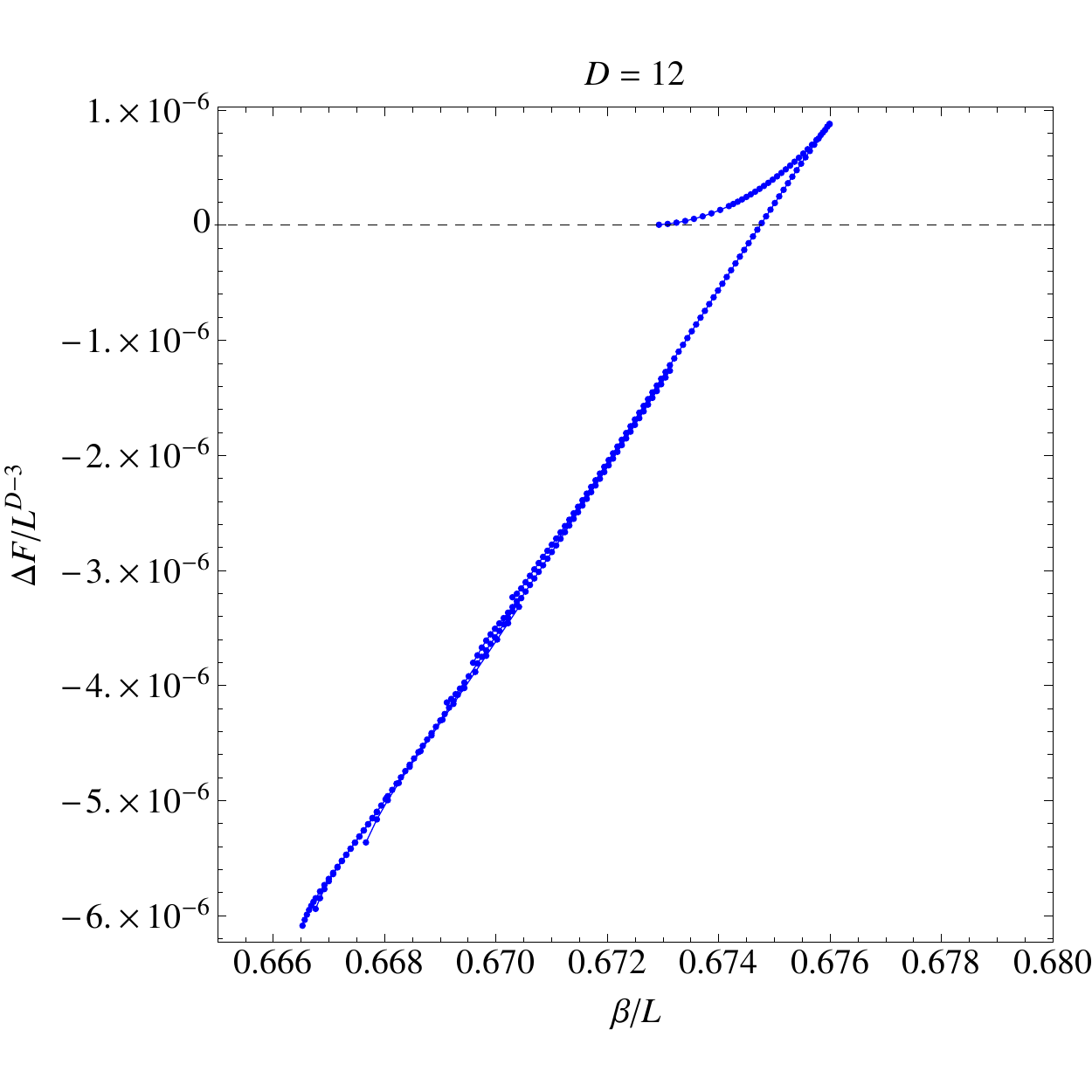}
\includegraphics[scale=0.65]{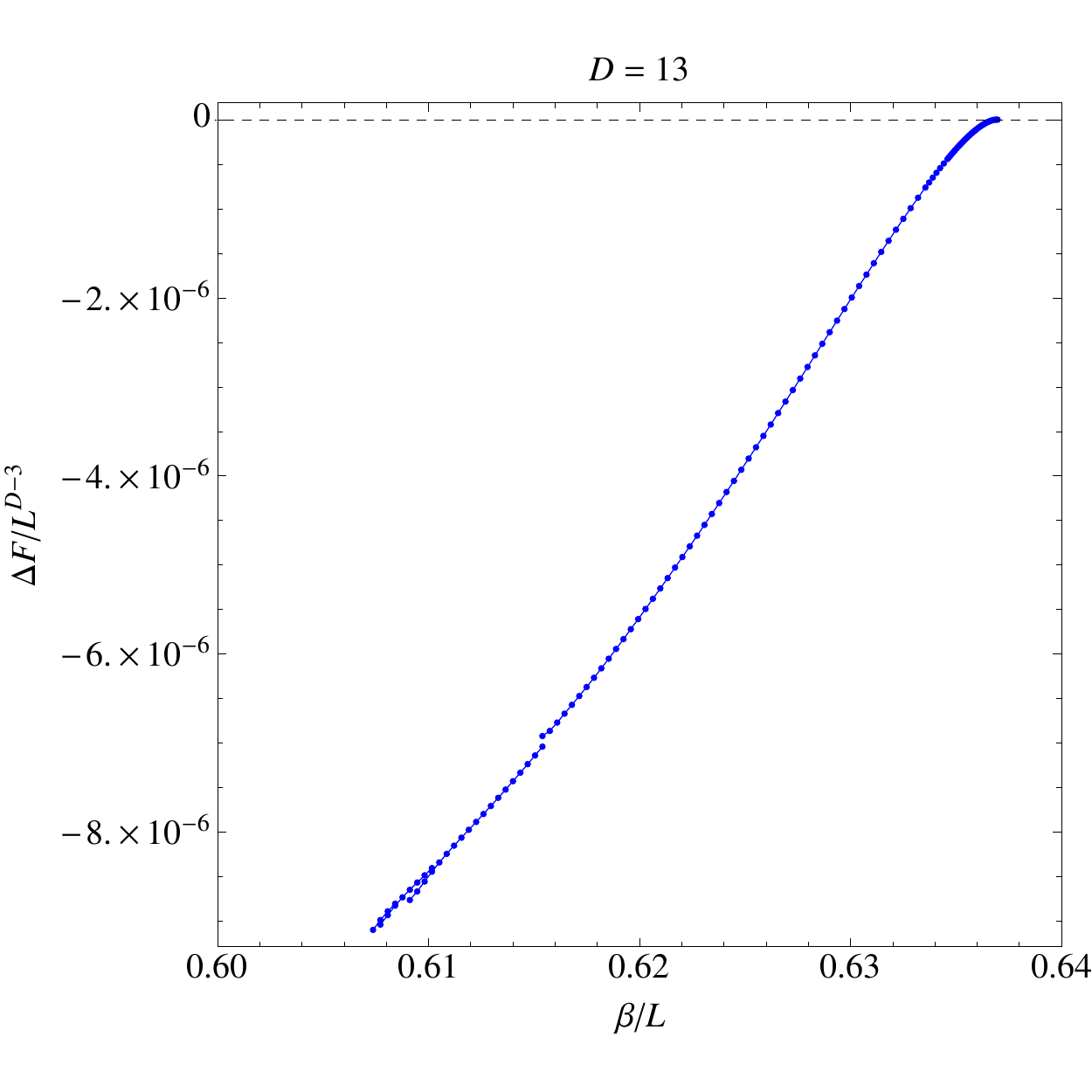}
\end{center}
\caption{Difference between the free energy of NUBSs and uniform black strings for the same temperature. The multiple curves correspond to data at different resolutions.  In $D=11$ NUBSs always have greater free energy than uniform strings with the same temperature. In $D=12$, $\Delta F$ reaches a maximum at the minimum temperature and it becomes negative for $\beta\leq 0.6748\,L$, indicating that for these temperatures the NUBSs dominate the canonical ensemble. In $D\geq 13$ the NUBS always have a lower free energy than the uniform black strings with the same temperature. }
\label{fig:freeenergy}
\end{figure}

In Figure \ref{fig:freeenergy} we plot, for various $D$, $\Delta F = F - F^0$ where $F$ is the free energy of a NUBS and $F^0$ the free energy of a uniform black string with the same temperature. 

For $D=11$, the NUBSs always have greater free energy and therefore they never dominate the canonical ensemble. 

For $D=12$ the free energy exhibits a cuspy behaviour. Small $\lambda$ NUBSs have greater free energy than uniform black strings. As $\lambda$ increases, $\Delta F$ reaches a maximum at the minimum temperature $\lambda=\lambda_0$, i.e., $\beta =  \beta_0 = 0.6760\,L$.\footnote{To see why these extrema must coincide, note that
$dF/d\lambda = dM/d\lambda-TdS/d\lambda - S dT/d\lambda$. The first law implies that this vanishes at $\lambda=\lambda_0$ since $dT/d\lambda=0$. Hence $F$ is extremized at $\lambda=\lambda_0$. Furthermore, $F^0$ is a single-valued function of $T$ so
$dF^0/d\lambda = (dF^0/dT) dT/d\lambda$, which also vanishes at $\lambda=\lambda_0$. Hence, viewed as a function of $\lambda$, $F^0$ also is extremized at $\lambda_0$. Therefore $\Delta F$ also is extremized.} 
For $\lambda>\lambda_0$, $\Delta F$ decreases, eventually becoming negative. It becomes zero at $\beta = 0.6748\,L$, so NUBSs with smaller $\beta$ are preferred over uniform strings in the canonical ensemble.

For $D= 13,14,15$ the NUBSs always have lower free energy than the uniform strings.  

\subsubsection{Negative modes}

\begin{figure}[t!]
\begin{center}
\includegraphics[scale=0.65]{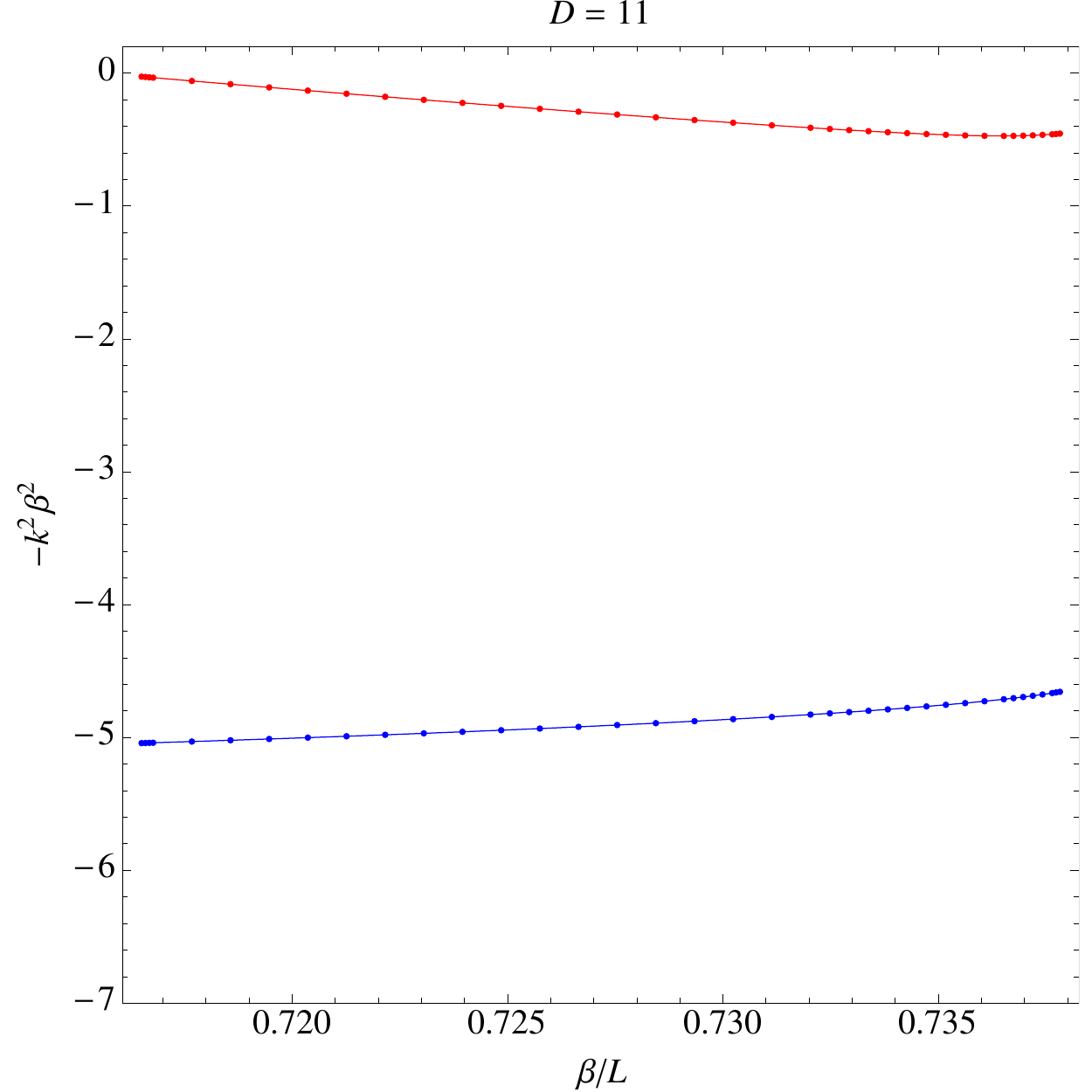}
\includegraphics[scale=0.65]{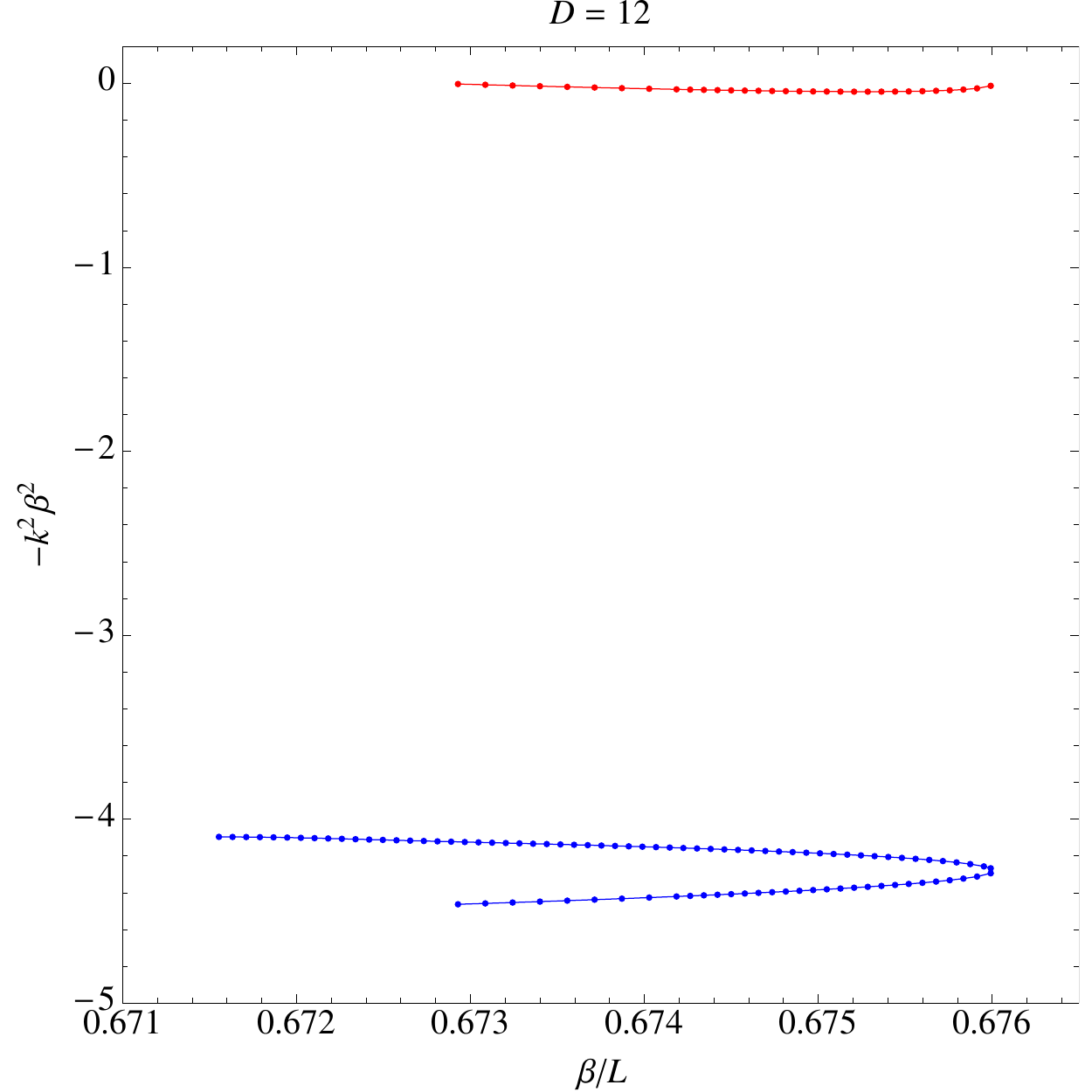}
\includegraphics[scale=0.65]{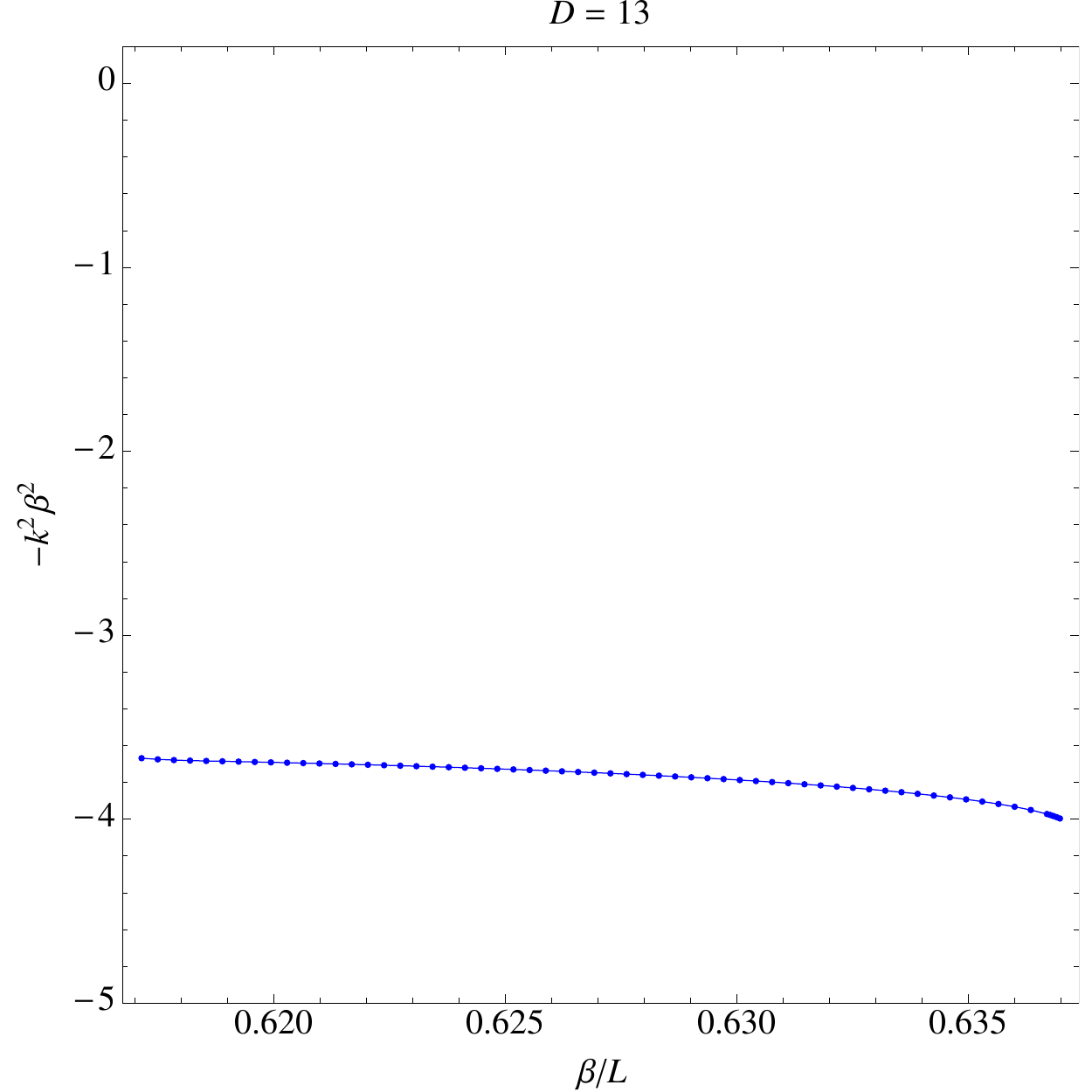}
\end{center}
\caption{Negative modes of the Lichnerowicz operator $\Delta_L$ in various dimensions. For $D=11$ NUBSs have two negative modes all along the branch. For $D=12$, weakly non-uniform strings (i.e., small $\lambda$) have two negative modes; at the minimum temperature ($\beta=\beta_\textrm{max}$) the mode with the least negative eigenvalue turns into a zero mode. For larger values of $\lambda$ there is only one negative mode. For $D\geq 13$ NUBSs have only one negative mode. }
\label{fig:negativemodes}
\end{figure}

Next we consider the spectrum of the Lichnerowicz operator $\Delta_L$ restricted to modes which respect the $U(1)_\tau\times SO(D-2)$ isometries of the background. We will only be interested in the negative modes of $\Delta_L$, which are an invariant  of the manifold in question. Our results for different $D$ are displayed in Figure \ref{fig:negativemodes}.

For $D=11$ we find that NUBSs have two negative modes, at least for all values of $\lambda$ that we have been able to explore. The mode with the most negative eigenvalue is continuously connected (as $\lambda \rightarrow 0$) to the uniform black string negative mode arising from the Euclidean Schwarzschild negative mode. The other NUBS negative mode reduces to the GL zero mode as $\lambda \rightarrow 0$. It is interesting to note that there are no other zero modes along the NUBS branch, just as in  $D=5$ \cite{Headrick:2009pv}. However, we do not see that the most negative mode becomes increasingly negative along the branch; in fact, both negative modes appear to remain finite as $\lambda$ increases. Of course, we have only been able to construct solutions up to moderate values of $\lambda$ and  it is entirely plausible that the mode with the most negative eigenvalue diverges at the merger point, as in $D=5$ \cite{Headrick:2009pv}.  

For $D=12$ we find that NUBSs with $\lambda<\lambda_0$ have two negative modes, exactly as in the lower dimensional cases. However, one of the negative modes reduces to a zero mode at $\lambda=\lambda_0$ and is absent for $\lambda>\lambda_0$. The zero mode at $\lambda=\lambda_0$ is obtained by an infinitesimal variation of parameters of the NUBS solution. Only at an extremum of temperature does such a variation of parameters respect the boundary conditions imposed on a Euclidean negative mode, namely that it should preserve the temperature. Hence a Euclidean negative mode can reduce to a zero mode corresponding to a variation of parameters in the background solution only at an extremum of temperature  \cite{Dias:2010eu}. For $\lambda>\lambda_0$ we find that there is only one negative mode and no other zero modes. 

For $D\geq 13$, we find that all NUBSs have one, and only one, negative mode. This mode is continuously connected (as $\lambda \rightarrow 0$) to the negative mode of the uniform black string inherited from the Schwarzschild solution.

\begin{figure}[t!]
\begin{center}
\includegraphics[scale=0.45]{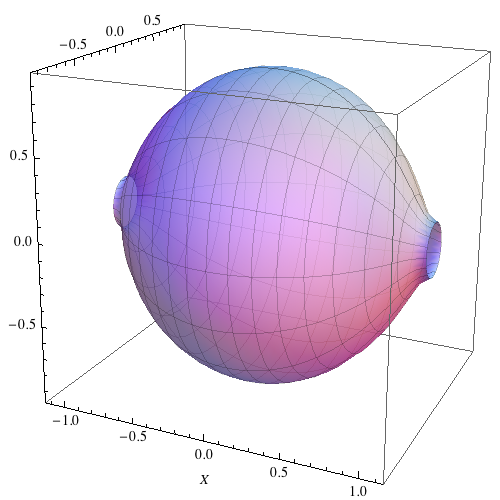}
\hspace{0.1cm}
\includegraphics[scale=0.55]{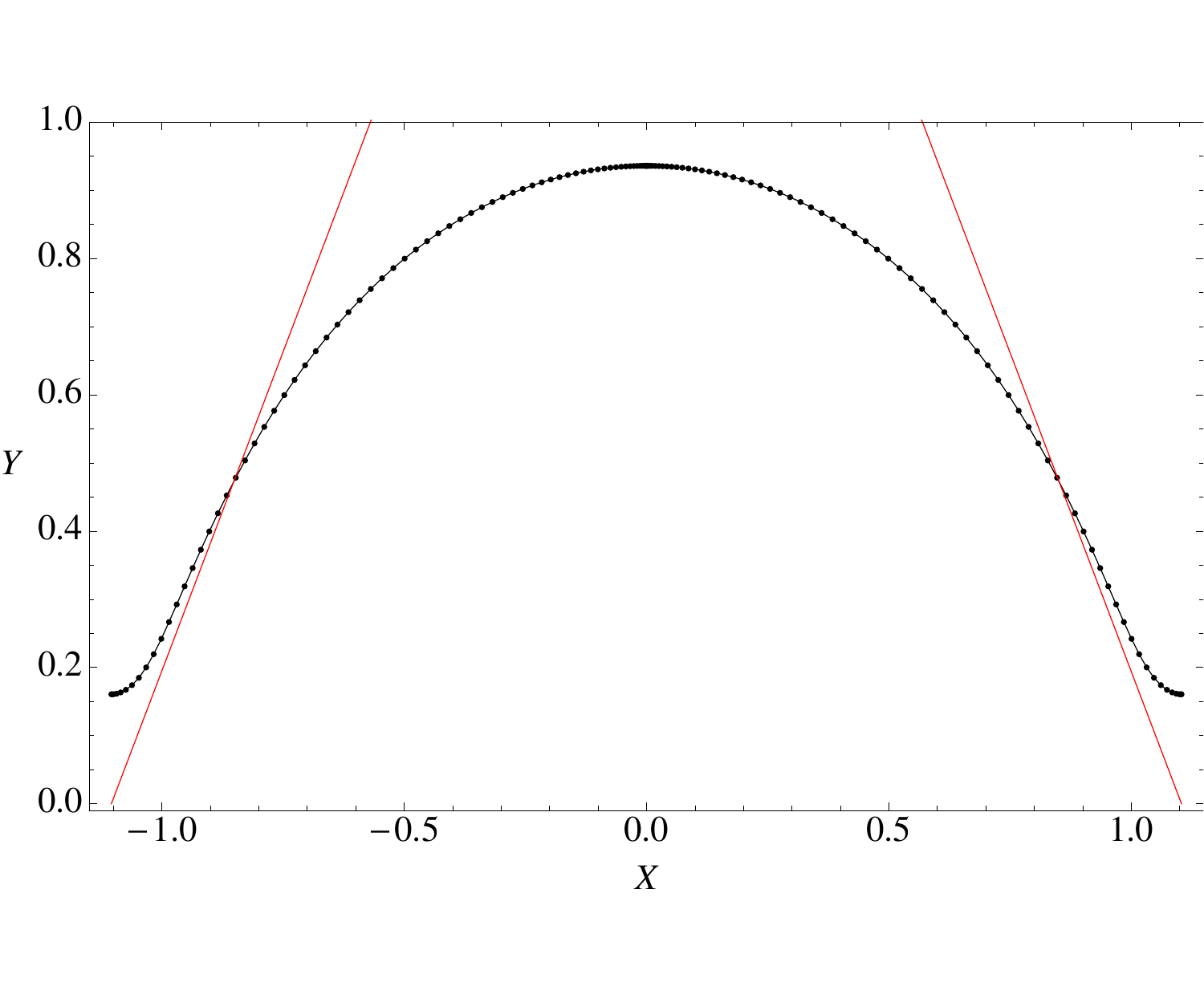}
\end{center}
\caption{Embedding of the horizon geometry of a $D=11$ NUBS with $\lambda = 2.41$ into $\mathbb R^{D-2}$. On the left figure we have plotted the radius vs. $X$ as a surface of revolution. On the right we compare the embedded geometry with that of a Ricci flat cone. The geometry starts approximating the Ricci flat cone metric of Ref. \cite{Kol:2002xz}.}
\label{fig:conegeometry}
\end{figure}

\subsubsection{Local geometry}

Finally we consider the local geometry near the minimum radius $S^{D-3}$
of the horizon. Recall that Ref. \cite{Kol:2002xz} predicted that
precisely at the merger point, the local geometry should be given by
that of a cone over $S^2\times S^{D-3}$, where first sphere comes
fibering the $U(1)_\tau$ and the second sphere comes from the
axisymmetry of the problem. Therefore, one would expect that the local
geometry near the minimal sphere of a NUBSs approximates that of a cone
only if the non-uniformity parameter $\lambda$ is sufficiently
large. Following Refs. \cite{Kol:2003ja,Sorkin:2006wp}, in Figure
\ref{fig:conegeometry} we have embedded the geometry of the horizon of a
$D=11$ NUBS with $\lambda = 2.41 $  into $\mathbb
R^{D-2}$
\begin{equation}
ds_{\mathbb R^{D-2}}^2=dX^2+dY^2+Y^2\,d\Omega_{D-3}^2\,,
\end{equation}
and projecting onto an equatorial circle of the $S^{D-3}$ in the plots. It is interesting to note that
as $D$ increases the geometry of a horizon of given $\lambda$ looks ``rounder'' (compare with the embedding plots in Refs. \cite{Wiseman:2002zc,Sorkin:2006wp}). 
To compare with the cone geometry we have also embedded the latter; as Figure \ref{fig:conegeometry} shows, the geometry of the horizon of the $\lambda = 2.41 $ NUBSs is reasonable close to that of the cone, but given that $\lambda$ is not large one should not expect  very good agreement.  For the solutions that we have constructed with $D>11$,  $\lambda$ is even smaller and the geometry  near the minimal horizon sphere is still quite far from the cone. 

\section{Local Penrose inequality}
\label{LPI}

\subsection{Introduction}

To investigate stability, it is convenient to label NUBSs by their mass $M$ instead of $\lambda$. For those $D$ in which the mass exhibits a maximum at $\lambda=\lambda_1$, we divide NUBSs into two families: "thin" non-uniform strings with $0<\lambda<\lambda_1$, and "fat" non-uniform strings with $\lambda>\lambda_1$. Solutions belonging to each family are uniquely labelled by $M$. Stability requires that an arbitrarily small perturbation leads to an arbitrarily small change in the solution, so a stable thin (fat)  solution cannot evolve to a fat (thin) solution when perturbed. Hence, in investigating stability we can regard thin and fat strings as distinct families of solutions. Let $A_{\rm NUBS}(M)$ denote the horizon area of the particular family of interest.

We will study stability using the method of local Penrose inequalities introduced in Ref. \cite{Figueras:2011he}. Consider a 1-parameter family of time-symmetric initial data for the Einstein equation, specified by a parameter $\epsilon$, with the following properties: (a) it is asymptotically flat in the KK sense, with the KK circle having circumference $L$ at infinity;  (b) for $\epsilon=0$ it reduces to the initial data on a constant $\tau$ slice (i.e. the Einstein-Rosen bridge) of a NUBS with mass $M_i$. For small $\epsilon$, our initial data describes a perturbation of the NUBS with mass $M_i$. Assume that this solution is stable. This means that, for sufficiently small $\epsilon$, the perturbation will disperse through radiation to infinity and across the horizon, and the spacetime will settle down to a new stationary solution, with mass $M_f(\epsilon)$, close to $M_i$. In general, the final state might have some angular or linear momentum and would belong to a stationary generalization of the static NUBS family of solutions. Since we know nothing about this generalization, we can exclude this possibility by imposing the addition restrictions: (c) the initial data preserves the $SO(D-2)$ symmetry of our NUBS solutions; (d) the initial data has a reflection symmetry around the KK circle. Our final state must then belong to the same family (thin or fat) of static NUBS solutions as our initial solution.

Let $A_{\rm app}(\epsilon)$ denote the area of the apparent horizon of the initial data, and $A_i(\epsilon)$ the area of the intersection of the event horizon with the initial data surface. The apparent horizon lies within the event horizon.\footnote{This assumes that cosmic censorship is not violated in the evolution, which is justified because a violation of cosmic censorship would be an instability, contradicting our assumption of stability.} The time symmetry of our initial data implies that the apparent horizon is a stable minimal surface and hence has smaller area than any surface which encloses it. Hence
\be
 A_{\rm app}(\epsilon) \le A_i(\epsilon)  \le A_{\rm NUBS}(M_f(\epsilon)) \le A_{\rm NUBS}(M(\epsilon))
\ee
The second inequality follows from the Hawking area theorem. In the third inequality, we use $M_f(\epsilon) \le M(\epsilon)$ where $M(\epsilon)$ is the mass of our initial data. This follows from the decrease of Bondi energy (i.e. gravitational waves carry away energy). The above inequality is referred to as the local Penrose inequality for the NUBS. The word "local" refers to the fact that it should hold only for arbitrarily small $\epsilon$. It is a necessary condition for stability of the NUBS. 

This inequality is saturated by initial data corresponding to the NUBS solution. Hence a stable NUBS is a local maximum of horizon area in the space of initial data of given mass, and satisfying the above conditions.\footnote{
Conditions (c) and (d) are not required to reach this conclusion: they can be eliminated if one knows that a generalization of the static NUBS with non-vanishing linear or angular momentum has lower horizon area than a static NUBS of the same mass. This follows from the first law and the assumption that the linear and angular velocities of the horizon have the same sign as the linear and angular momentum respectively.} The first law implies that an unstable NUBS is an extremum of horizon area at fixed mass, but this extremum is not a local maximum. 

We are interested in arbitrarily small $\epsilon$ so we expand both sides of this inequality in $\epsilon$. At zeroth order we have $M(0)=M_i$, $A_{\rm app}(0)=A_{\rm NUBS}(M_i)$ and so the inequality is saturated. At first order the inequality remains saturated because of the first law of black hole mechanics $\dot{M}(0) = (T/4) \dot{A}_{\rm app}(0)$ where a dot denotes a derivative with respect to $\epsilon$ and $T$ is the Hawking temperature of the NUBS with mass $M_i$. At second order in $\epsilon$, the local Penrose inequality reduces to
\be
\label{localPenroseinequality}
 Q \ge 0, \qquad Q \equiv \ddot{M}(0) - \frac{1}{4} T \ddot{A}_{\rm app}(0) - \frac{1}{T c_L} \dot{M}(0)^2
\ee
where $c_L = (dM/dT)_L$ denotes the heat capacity of the NUBS (with $L$ held fixed).

To summarize, a necessary condition for stability of the NUBS is that initial data describing a small perturbation of the NUBS must satisfy the inequality (\ref{localPenroseinequality}). The definition of $Q$ involves quantities which depend on the perturbation to second order in $\epsilon$. However, the constraint equations governing second order perturbations of initial data reveal that such perturbations are sourced by quantities quadratic in first order perturbations and it can be argued using the first law that $Q$ itself depends only on the first order perturbation \cite{Figueras:2011he}.

Is the condition (\ref{localPenroseinequality}) also sufficient for stability? Ref. \cite{Hollands:2012sf} argued that it is both necessary and sufficient for the stability of linearized perturbations. This was done by defining a quantity ${\cal E}$ that can be expressed as an integral over a spacelike surface of a quantity quadratic in a linearized perturbation, and showing that ${\cal E}$ is a non-increasing function of time. If all perturbations have ${\cal E} \ge 0$ then linearized perturbations cannot grow, strongly suggesting stability. However, if there exists a perturbation with ${\cal E}<0$ then such a perturbation cannot decay, strongly suggesting instability. Ref. \cite{Hollands:2012sf} proved that there exists a perturbation for which ${\cal E}<0$ if, and only if, there exists a perturbation for which $Q<0$. 

Let's apply the above arguments to a NUBS solution with $D \le 13$. Consider first the solution corresponding to the intersection of the uniform and non-uniform families of solutions.  We can regard this as a NUBS with $\lambda=0$. Consider the perturbation of this solution corresponding to an increase in mass, moving along the uniform string branch. This gives a uniform string with a greater horizon area than a NUBS with the same mass. Hence this perturbation violates the local Penrose inequality for the NUBS family. So if we regard this solution as a member of the NUBS family then it is unstable. Now consider a NUBS with very small non-zero $\lambda$. Construct initial data describing a perturbation of such a solution, such that it reduces to the perturbation just discussed in the limit $\lambda \rightarrow 0$. Expanding $Q$ in powers of $\lambda$ gives\footnote{One can show $Q_1=0$ but that doesn't matter here.} $Q =Q_0 + \lambda Q_1 + \lambda^2 Q_2 + \ldots$. We've just seen that $Q_0<0$. Hence for sufficiently small positive $\lambda$ we will have $Q<0$. Therefore a $D \le 13$ NUBS with small positive $\lambda$ will be unstable.

We have seen that, for $D=12,13$, there exists a maximum mass along the branch of NUBS solutions. The same occurs in other contexts, e.g. rotating neutron stars or black rings at fixed angular momentum $J$. In these other systems, a heuristic "turning point" argument suggests that an unstable mode appears when one passes through the maximum of mass, on the side in which the heat capacity is positive \cite{Friedman:1988er,Arcioni:2004ww}. The local Penrose inequality provides a more rigorous justification for this argument \cite{Figueras:2011he}. At a maximum of mass $c_L$ ($c_J$ in the case of rotating solutions) changes sign by passing through zero. Hence the final term of (\ref{localPenroseinequality}) diverges, and is negative on the side in which $c_L$ is positive. This strongly suggests that solutions close to the maximum mass solution, with positive $c_L$ should be unstable. Note that it does {\it not} suggest any change in stability at the minimum temperature solutions, where $c_L$ changes sign by passing through infinity (this is why there is no instability of the Kerr solution when the heat capacity at fixed angular momentum changes sign \cite{Sorkin:1982ut}).

Our $D=12,13$ NUBS solutions exhibit a maximum mass at $\lambda=\lambda_1$, with $c_L$ positive for $\lambda<\lambda_1$, so the above argument strongly suggests instability for $\lambda$ close to, but less than, $\lambda_1$. 

\subsection{Instability of non-uniform black strings}
\label{subsec:instability}

We have argued above that NUBSs are unstable if $D \le 13$ and $\lambda$ is infinitesimal, or if $D=12,13$ and $\lambda$ is slightly less than $\lambda_1$. In this section we will use the local Penrose inequality to investigate the stability of NUBSs with other values of $D$ and $\lambda$. 

\subsubsection{Simple initial data}
\label{subsubsec:simpleinitdat}

Let $\bar{h}_{ab}$ the induced metric on a surface of constant $\tau$ in a NUBS solution. The extrinsic curvature of such a surface vanishes, so we  consider  time-symmetric initial data and hence the momentum constraint is trivially satisfied. Construct new time symmetric initial data, describing a perturbation of this solution, by conformally rescaling the induced metric:
\begin{equation}
\label{ID}
h_{ab}=\Psi^{4/(D-3)}\bar{h}_{ab},\quad K_{ab}=0\ . 
\end{equation}
The Hamiltonian constraint then becomes
\begin{equation}
\bar{\nabla}^2 \Psi=0\,
\end{equation}
where $\bar{\nabla}$ is the covariant derivative with
respect to $\bar{h}_{ab}$. 
Expanding $\Psi$ in terms of a small parameter $\epsilon$ as 
$\Psi=1+\epsilon\, \dot{\Psi}(0)+\epsilon^2\,\ddot{\Psi}(0)/2\ldots$,\footnote{Here the dots $\dot\,$ denote derivatives with respect to the parameter $\epsilon$.} 
we can construct, perturbatively, initial data by solving
\begin{equation}
\label{eqpsi}
 \bar{\nabla}^2 \dot{\Psi}=\bar{\nabla}^2 \ddot{\Psi}=0\ .
\end{equation}

As explained in Ref. \cite{Figueras:2011he},
the  homogeneous solution for $\ddot{\Psi}$ does not contribute to the local
Penrose inequality. Therefore we choose the trivial solution $\ddot{\Psi}=0$.
For the first order solution $\dot{\Psi}$, 
we demand that the resulting induced metric is asymptotically KK, which amounts to impose $\dot{\Psi}=\mathcal{O}(1/r^{n-1})=\mathcal{O}(1-y)$ 
near infinity. With this decay at infinity, the asymptotic length of the KK in the perturbed space-time is the same as that of the NUBS background. 
In addition, we place the inner boundary at $y=y_b<0$ and 
impose Dirichlet boundary conditions there, i.e.,
we specify $\dot{\Psi}|_{y=y_b}$.  
We can do this because in our numerical solutions for the background NUBSs  all functions are even in $y$ so we can analytically continue the range of $y$ to negative values. Then, $y=0$ is the bifurcation surface and the region of the space-time with $y<0$ corresponds to another asymptotically KK region.   
Finally, we impose reflecting boundary conditions on $\Psi$ on both the reflection plane $(x=0)$ and on the periodic boundary $(x=1)$.
Solving the equation for $\dot{\Psi}$ with these boundary conditions,
we can obtain the variation of mass, the variation of the area of apparent horizon 
and finally $Q$.
We describe the details of the calculation of $Q$ in Appendix \ref{ELPI}.

For the results presented in Fig. \ref{Qmin}  we have imposed the simple boundary condition $\dot\Psi|_{y=y_b}=1$, and this turns out to be sufficient to detect instabilities for all values of $\lambda$ that we could explore. We also considered more general boundary conditions $\dot\Psi|_{y=y_b}=1+\sum_{n=1}^N c_n \cos(n \pi x)$ in order to look for instabilities for those values of $\lambda$ for which $Q>0$ with the previous simple boundary condition. Here 
the integer $N$ represents a truncation of the
Fourier expansion and we fixed the normalization of $\dot{\Psi}$ 
by choosing the first term in the
expansion to be unity. To look for initial data which satisfies $Q < 0$, 
we seek to determine the $\{c_n\}$ which minimize the dimensionless
quantity
\begin{equation}
 \bar{Q}=\frac{Q}{D^2 M}\ , \qquad D^2=\int^{L/2}_0 dx \,
  \dot{\Psi}(y=0,x)^2\ .
\end{equation}
We varied the number of Fourier coefficients $N$ in the previous expansions, but the qualitative results presented in Fig. \ref{Qmin} do not change.  Finally, we put the inner boundary at $y_b=-1.25\times 10^{-3}$, but have checked that our results do not depend on the specific choice of $y_b$.

To proceed with the calculation of $Q$, we first interpolate the (numerical) background solution using a Lagrange polynomial and on top of this background we consider the perturbations as we described above. In this calculation,  it is crucial to ensure that the numerical error of the background does not affect the value of $Q$. To do so, recall from the discussion in \S\ref{NumericalNUBS} that $\phi\equiv\xi_a\xi^a$ provides a measure of the numerical error of the background. Therefore, we have to estimate the value of $\phi=\phi_{acceptable}$ such that for $\phi<\phi_{acceptable}$ the value of $Q$ is independent of $\phi$. We have made this estimate by comparing the values of $Q$ obtained for  backgrounds  computed at two different resolutions (and hence different values of $\phi$) and we found that for $\phi_\textrm{max}<10^{-3}$ the value of $Q$ changes by less than 10\% as the resolution of the background changes. This estimate is independent of the number of space-time dimensions. Here $\phi_\textrm{max}$ denotes the maximum value of $\phi$ over the whole computational domain.  The calculation of $Q$ has its own numerical error, which we estimate by monitoring $\delta\equiv |\dot{M}-T\dot{A}_\textrm{app}/4|/(D M)$, which should
vanish by the first law of black hole mechanics. For all results
presented the error in the background turns out to be the dominant one and  $\delta<0.01$ in all cases.

Our results are presented in  Figure \ref{Qmin},  where we plot $\bar Q$ against the
non-uniformity parameter $\lambda$ in various dimensions.   For $D=11$, we see that $\bar Q$ is 
negative for any value of $\lambda$ for which we have acceptable numerical errors. Thus all $D = 11$ NUBSs we
have checked are unstable, even well in the non-perturbative regime (i.e., large $\lambda$). This is one of the main results of this paper.\footnote{Ref. \cite{Horowitz:2011cq} argued that slightly NUBS should be unstable by continuity.}  We have also checked that NUBS in $D=5,6,8,10$ are also unstable for different values of $\lambda$, which are not necessarily small. Therefore, it is reasonable to expect that all non-uniform black
strings in $5 \leq  D\leq 11$ are dynamically unstable. 

For $D=12,13$,  we find that 
$Q$ is negative for $\lambda<\lambda_1$, which implies that
all ``thin'' non-uniform strings in $D=12,13$ are unstable. Note the
divergence in $Q$ at $\lambda=\lambda_1$, where $c_L$ vanishes.\footnote{A similar situation was found in Ref. \cite{Figueras:2011he} for black rings: ``fat" rings were found to be unstable and ``thin" rings were found to be stable and $Q$ goes through infinity when moving from one branch to the other because the specific heat at constant angular momentum vanishes at the maximum area (at fixed mass) solution.} Thus our
results confirm the existence of an instability when $\lambda$ is close
to, but less than, $\lambda_1$. More importantly, they demonstrate that
an instability is present for all $\lambda<\lambda_1$.  For
$\lambda>\lambda_1$, we minimised $Q$ as explained above (with $N=3,4$) and we always found that $Q$ is small and positive and hence, there
is no indication of an instability in ``fat'' non-uniform strings. Given
the ease with which we found an instability for thin strings, this
strongly suggests that fat black strings are stable.  

Finally, for $D=14$ we found that after minimisation $Q$ is positive for any $\lambda$,  which again strongly
suggests that these solutions are stable. We  found the same result for
$D=15$.

\begin{figure}[t!]
  \centering
  \subfigure[$D=11$]
  {\includegraphics[scale=0.52]{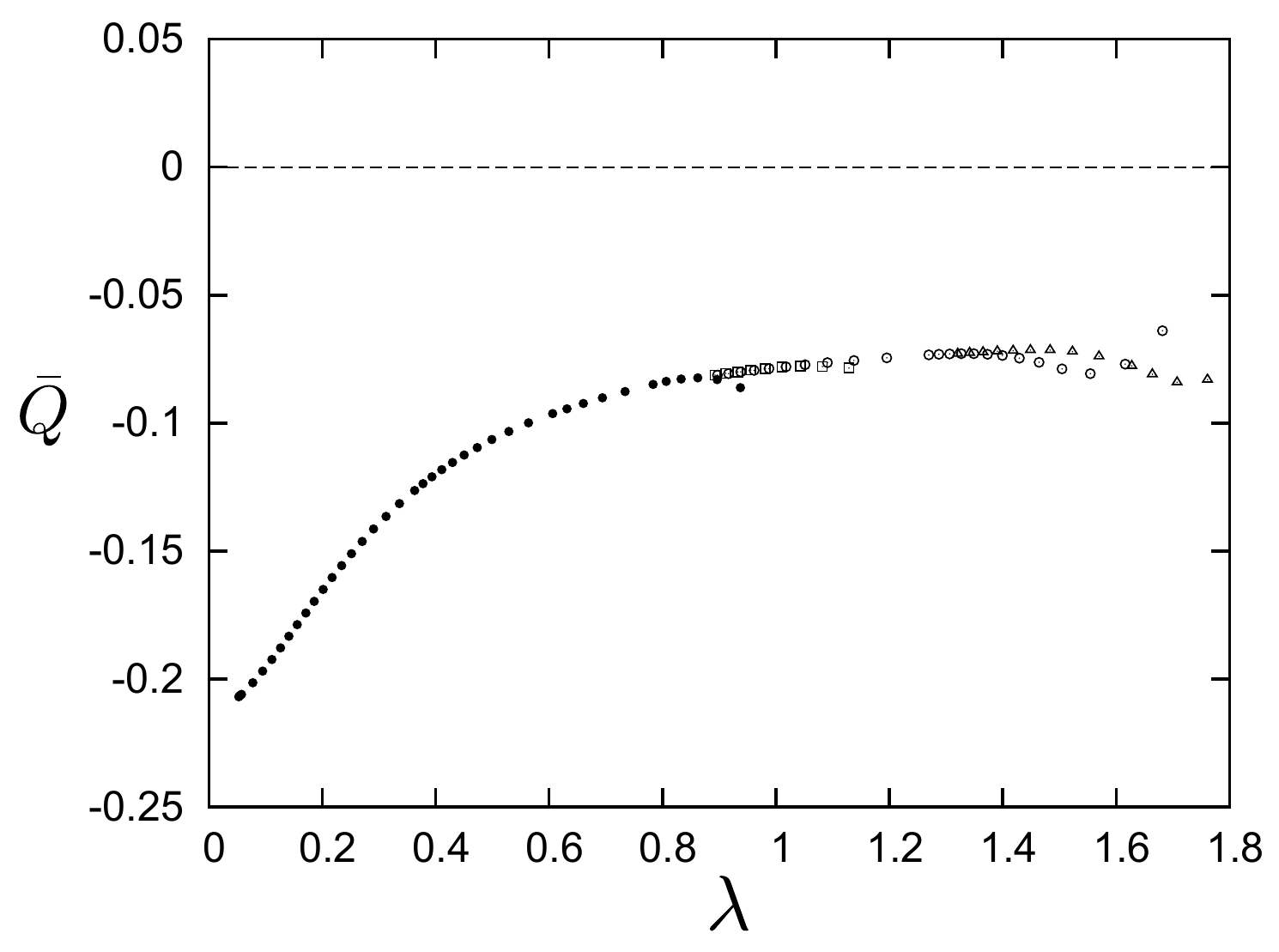}
  }
  \subfigure[$D=12$]
  {\includegraphics[scale=0.52]{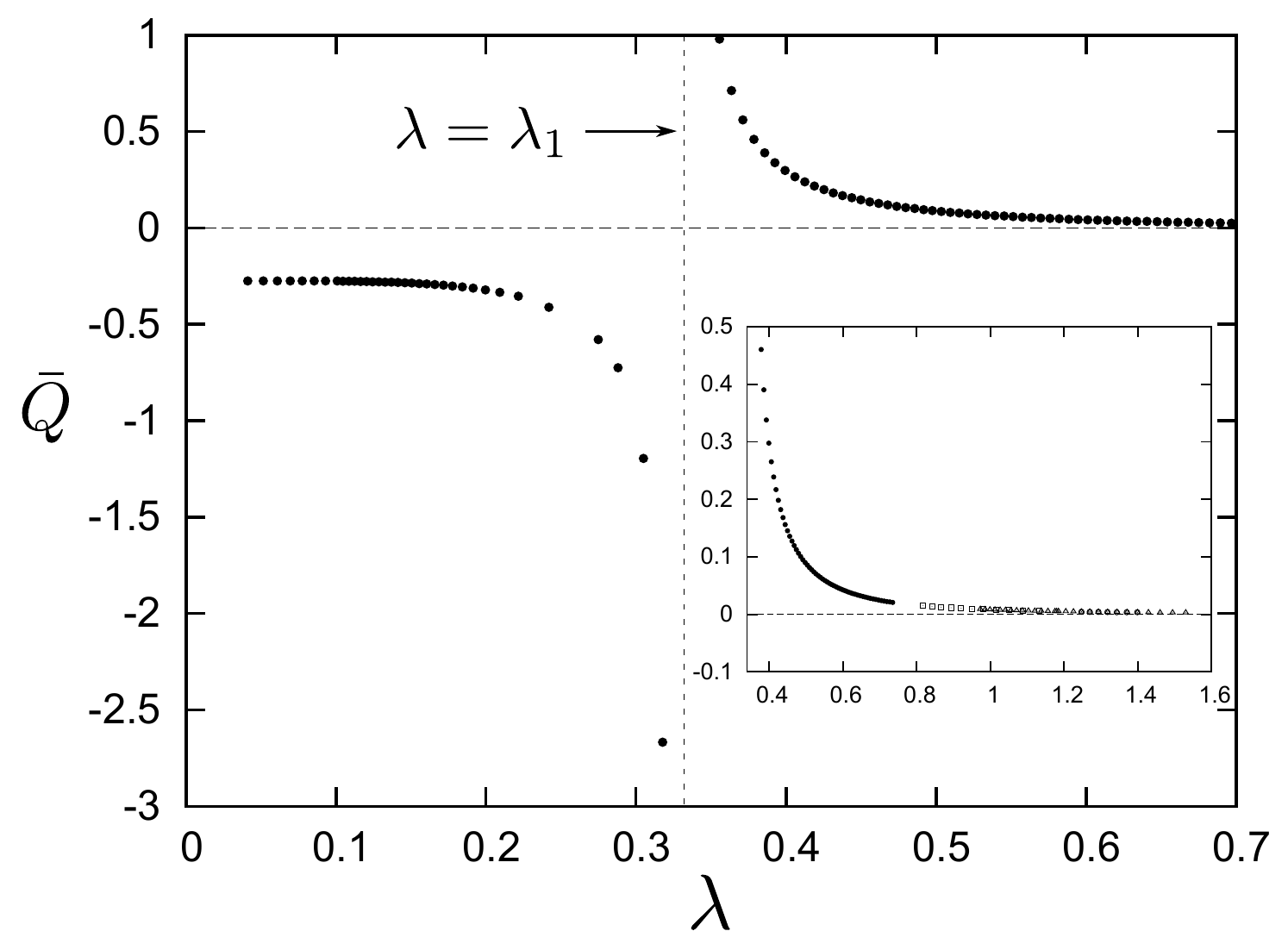} 
  }
  \subfigure[$D=13$]
  {\includegraphics[scale=0.52]{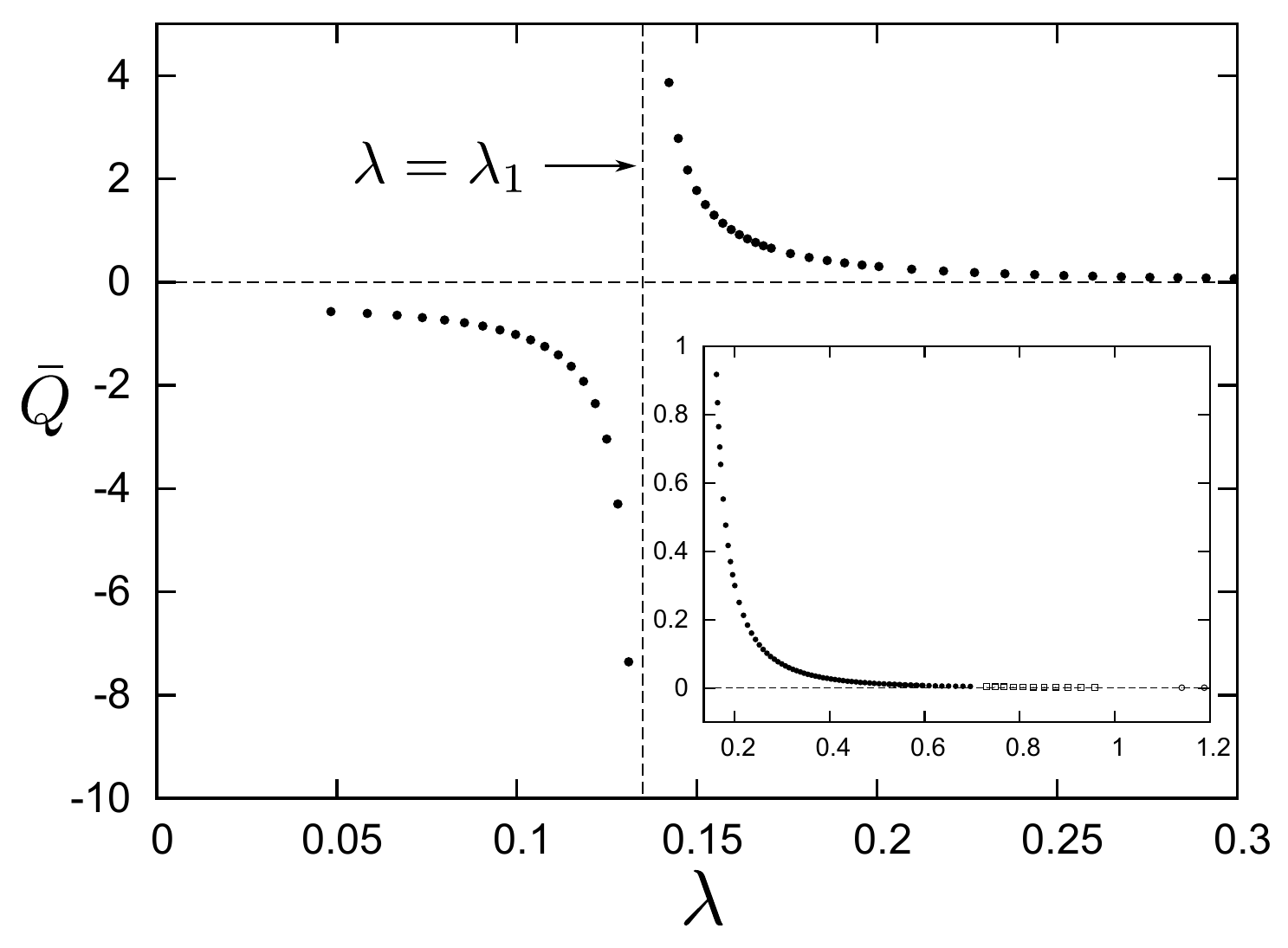}
  }
  \subfigure[$D=14$]
  {\includegraphics[scale=0.52]{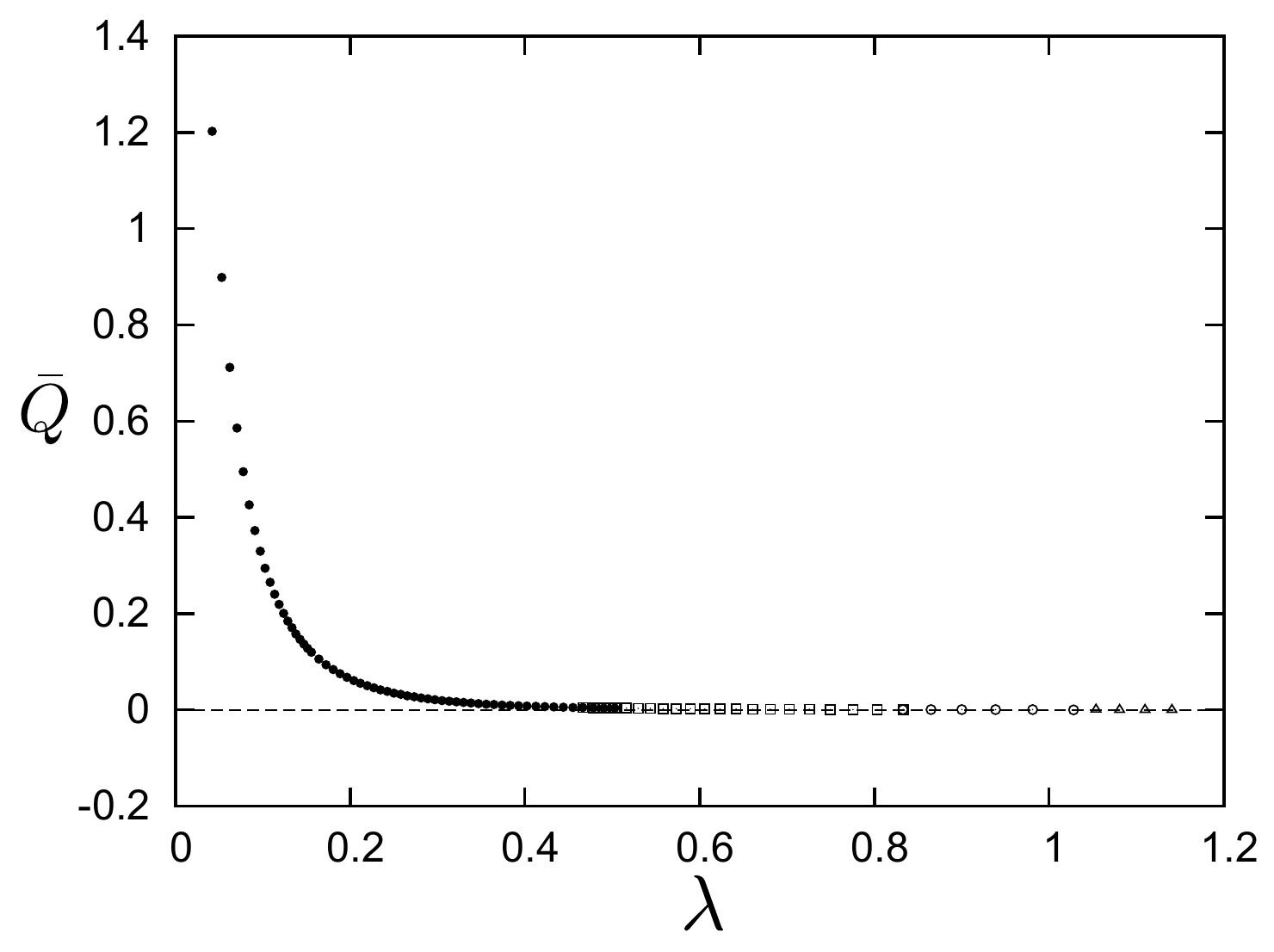}
  }
  \caption{
$\bar Q$ vs. $\lambda$ for $D=11,12,13,14$. 
In Figures (b) and (c)   ($D=12,13$ respectively) we have included insets showing a zoom in of the region with $\lambda>\lambda_1$ (``fat" NUBSs).  For $D=11$, $\bar Q$ is always negative, which implies these that NUBSs are unstable.
For $D=12$ and $13$, $\bar Q$ is negative for
 $\lambda<\lambda_1$ and positive for $\lambda>\lambda_1$.
Thus, ``thin'' NUBSs are unstable, but for ``fat'' NUBSs, there is no
 indication of instability.
For $D=14$, $\bar Q$ is positive for any $\lambda$ and 
there is no indication of instability.
\label{Qmin}
}
\end{figure}

\subsubsection{More general initial data}

We have also considered a more general class of time-symmetric initial data defined as follows. Instead of taking $\bar{h}_{ab}$ to be the induced metric on a surface of constant $\tau$ in the NUBS spacetime, consider the deformed metric
\begin{equation}
\bar h_{ab}\rightarrow 
\hat{h}_{ab}\,dx^adx^b=\frac{r_0^2\,e^{S}}{f(y)^{\frac{2}{D-4}}}\,e^{-2\Phi/(D-3)}\,d\Omega_{(D-3)}^2+e^{\Phi}\bigg[e^{A}\,dx^2+\frac{4\,r_0^2\,\Delta\,e^{B}}{f(y)^{\frac{2(D-3)}{D-4}}}\,(dy+y\,f(y)\,F\,dx)^2\bigg]\,,
\label{eqn:defmetric}
\end{equation}
where the function $\Phi(x,y)$ describes the deformation: 
for $\Phi=0$ the metric \eqref{eqn:defmetric} reduces to the induced metric of a NUBS.  Note that for an arbitrary $\Phi$ the induced Ricci scalar $\hat R$ is non-vanishing because the Hamiltonian constraint is not satisfied. 

To construct initial data that satisfies the constraints we proceed as before by conformally rescaling the induced metric as in \eqref{ID} and solving the Hamiltonian constraint in terms of the conformal factor:
\begin{equation}
\hat\nabla^2\Psi=\frac{D-3}{4(D-2)}\,\hat R\,\Psi\ ,
\label{eqn:defconstraint}
\end{equation}
where $\hat\nabla$ is the covariant derivative with respect to
$\hat{h}_{ab}$.
Note that $\Phi$ is a free function; once we have chosen $\Phi$ and specified boundary conditions, eq.\eqref{eqn:defconstraint} uniquely determines $\Psi$. The idea now is to expand both $\Psi$ and $\Phi$ in terms of the small parameter $\epsilon$. So as before we have $\Psi=1+\epsilon\, \dot{\Psi}(0)+\epsilon^2\,\ddot{\Psi}(0)/2+\ldots$, and we demand that $\Phi = {\cal O}(\epsilon)$ so that in the $\epsilon\to 0$ limit we recover the NUBS.  For simplicity (and because we can freely specify this function) we choose $\Phi = \epsilon\, \dot{\Phi}$,  and all the higher order terms in $\epsilon$ identically vanishing.  Then, expanding \eqref{eqn:defconstraint} in powers of $\epsilon$ up to second order gives 
\begin{equation}
\begin{aligned}
&\bar\nabla^2\dot\Psi=\frac{D-3}{4(D-2)}\,\dot{\hat R}\,,\\
&\bar\nabla^2\ddot\Psi=\frac{D-3}{4(D-2)}\big[\ddot{\hat R}
+2\,\dot{\hat R}\,(\dot\Psi+\dot\Phi)\big]+2\,\bar h^{ab}(\partial_a\dot\Phi)\,(\partial_b\dot\Psi)\,.
\end{aligned}
\label{eqn:pertdefeqs}
\end{equation}

Solving \eqref{eqn:pertdefeqs} subject to boundary conditions as in the previous subsubsection we have constructed, perturbatively, initial data for different choices of $\dot\Phi$. We have proceeded as in \S\ref{subsubsec:simpleinitdat} and expanded the inner boundary condition for $\dot\Psi$ in a Fourier series and then minimised $Q$. Note that now $\ddot\Psi$ is sourced by $\dot\Phi$ and $\dot\Psi$, so even if we impose $\ddot\Psi|_{y=y_b}=0$ (which we have), this function will be non-vanishing.\footnote{As explained in Ref. \cite{Figueras:2011he}, changing the inner boundary condition for $\ddot\Psi$ amounts to modify the homogeneous solution to the equation which, by the first law, cannot affect the Penrose inequality.} Therefore, in general we have $\ddot M(0)\neq 0$. 

The results that we have obtained for the Penrose inequality in various dimensions using this more general class of initial do not qualitively differ from those reported in \S\ref{subsubsec:simpleinitdat}. Therefore we do not display them here. It is worth emphasising that no matter the choice of $\dot\Phi$, we always find that $Q$ is positive in $D=12,13$ for ``fat'' NUBSs and in $D=14,15$ for all NUBSs.  This is further evidence that such solutions ought to be stable.

\medskip

\section*{Acknowledgments}

We are grateful to Toby Wiseman for useful discussions. PF is supported by an EPSRC postdoctoral fellowship [EP/H027106/1].
KM is supported by JSPS Grant-in-Aid for Scientific Research No.24$\cdot$2337.
HSR is supported by a Royal Society University Research Fellowship and by European Research Council grant no. ERC-2011-StG 279363-HiDGR. Some of the computations presented were done on COSMOS at the UK National Cosmology Supercomputing Centre in Cambridge. 

\appendix
\section{Details of the calculation of the local Penrose inequality}
\label{ELPI}
In this appendix we give more details on the calculation of $Q$ (see
Eq.(\ref{localPenroseinequality}) for the definition) for the class of initial data in \S\ref{subsubsec:simpleinitdat}. 
To do so, we need to calculate the 
variation of the mass 
and area of apparent horizon 
induced by the perturbation.

Lets first recall the expressions for the mass in asymptotic
Kaluza-Klein spacetimes, i.e., the metric at infinity approaches
$M_{D-1}\times S^1$, where $M_{D-1}$ denotes the $(D-1)$-dimensional Minkowski
space. 
In the linearised regime one finds that the spatial components of the
metric, in a Cartesian coordinate system,  are given by
\cite{hep-th/0309116}:
\begin{equation}
h_{zz}=\frac{c_z}{r^{D-4}}\,,\qquad h_{ij}=\delta_{ij}\,\frac{c_t-c_z}{(D-4)\,r^{D-4}}\,,
\end{equation}
where $r=\sqrt{\sum_{i=1}^{D-2}(x^i)^2}$ is the Euclidean radial
distance, $z\sim z+L$ is the compact direction and $c_t$ and $c_z$ are
two constants. Then the mass is given by
\begin{equation}
M=\frac{L\,\Omega_{D-3}}{16\pi}\big[(D-3)c_t-c_z\big]\,,
\end{equation}
where $\Omega_{D-3}$ is the area of a $(D-3)$-dimensional sphere.
Given the form of our initial data, Eq.(\ref{ID}), we can easily
calculate the variation of 
these parameters due to the perturbation $\dot\Psi$:
\begin{equation}
\dot c_z=\frac{4}{D-3}(r^{D-4}\dot\Psi)_{r=\infty}\,,\qquad \dot
 c_t=(r^{D-4}\dot\Psi)_{r=\infty}\ .
\end{equation}
Since we set $\ddot\Psi$ to be zero, we have 
$\ddot c_z=\ddot c_t=0$. Thus, there is no second order
perturbation in variation of the mass.
The first order variation of mass due to our perturbations is
\begin{equation}
\label{dM}
\dot M=\frac{L\,\Omega_{D-3}}{4\pi}\,\frac{(D-3)^2-1}{D-3}
(r^{D-4}\dot\Psi)_{r=\infty}\,.
\end{equation}

Now, we consider the variation of area of apparent horizon. 
In Ref.\cite{Figueras:2011he}, 
it was shown
that the apparent horizon coincides with the minimal surface for
axisymmetric perturbations.
Thus, we calculate area of the minimal surface instead of that of
the apparent horizon.
We denote the location of the minimal surface as $y=Y(x)$.
Then, the induced metric on the minimal surface is 
\begin{equation}
\label{indm}
\gamma_{IJ}dx^I dx^J=\Psi^{4/(D-3)}\left[
\frac{r_0^2\,e^{S}}{f(Y)^{\frac{2}{D-4}}}\,d\Omega_{D-3}^2
+e^A dx^2
+\frac{4\,r_0^2\,\Delta\,e^{B}}{f(Y)^{\frac{2(D-3)}{D-4}}}
(Y'+Yf(Y)F)^2dx^2
\right]\ ,
\end{equation}
where $'\equiv d/dx$.
In the above expression, the functions $A$, $B$, $F$, $S$ and $\Psi$ 
are evaluated at 
$y=Y(x)$ and they depend only on $x$.
The area of the minimal surface is given by 
$A_\textrm{app}=\int d^{D-2}x\sqrt{\gamma}$ where 
$\gamma=\textrm{det}(\gamma_{IJ})$.
We expand $Y(x)$ and $A_\textrm{app}$ in term of $\epsilon$ as 
$Y(x)=\epsilon \dot{Y}(x)+\cdots$ and 
$A_\textrm{app}=A_\textrm{BG}+\epsilon \dot{A}_\textrm{app} + \epsilon^2\ddot{A}_\textrm{app}/2+\cdots$ 
where $A_\textrm{BG}$ is the area of the event horizon in the background geometry.
Then, we obtain first and second order
perturbation of the area of the minimal surface as
\begin{eqnarray}
 &&\dot{A}_\textrm{app}=\frac{2(D-2)}{D-3} \Omega_{D-3}\, r_0^{D-3} \int_H dx \,
e^{(A+(D-3)S)/2} \dot{\Psi}\ .\\
&&\ddot{A}_\textrm{app}
= \Omega_{D-3} r_0^{D-3} \int_H dx\,
e^{(A+(D-3)S)/2} 
\bigg[
4r_0^2\Delta e^{B-A}(\dot{Y}'+\dot{Y}F)^2
+\frac{2(D-3)}{D-4}\dot{Y}^2 \nonumber\\
&&\hspace*{2cm}
+\frac{1}{2}(\partial_y^2 A+(D-3)\partial_y^2 S)\dot{Y}^2
+\frac{4(D-2)}{D-3}\partial_y \dot{\Psi} \dot{Y}+\frac{2(D-1)(D-2)}{(D-3)^2}\dot{\Psi}^2
\bigg]
\end{eqnarray}
where functions $A,B,F,S,\dot{\Psi}$ and their derivatives are evaluated
at background horizon $y=0$.
Since $\dot{Y}(x)$ extremizes the area functional $\ddot{A}_\textrm{app}$,
we have the equation for $\dot{Y}(x)$ as
\begin{multline}
\label{eqAPP}
-8r_0^2\Delta e^{-(A+(D-3)S)/2}[e^{(-A+2B+(D-3)S)/2}(Y'+YF)']'\\
+8r_0^2\Delta e^{B-A}(Y'+YF)F
+(\partial_y^2 A+(D-3)\partial_y^2 S)\dot{Y}
+\frac{4(D-2)}{D-3}\partial_y \dot{\Psi}=0\ .
\end{multline}
Using this equation, we obtain a simple expression for
$\ddot{A}_\textrm{app}$ as
\begin{equation}
\label{ddA}
\ddot{A}_\textrm{app}
= \frac{2(D-2)}{D-3} \Omega_{D-3}\, r_0^{D-3} \int_H dx \,
e^{(A+(D-3)S)/2} 
\big[
\dot{Y} \partial_y \dot{\Psi}+\frac{D-1}{D-3}\dot{\Psi}^2
\big]\ .
\end{equation}

In the calculation of $Q$ we do not need the first order
variation of the area of the apparent horizon $\dot{A}_\textrm{app}$. However,  we use this quantity  
to check the first law of the black hole thermodynamics 
$\dot{M}-T\dot{A}_\textrm{app}/4=0$, which provides an estimate of the numerical error in our  calculations.

\end{document}